\documentclass[aps,pre,showpacs,
amssymb,twocolumn
]{revtex4}
\usepackage{graphicx}
\usepackage{amsmath}
\usepackage{amsfonts}
\usepackage{amssymb}
\usepackage{epsfig,libertine}
\usepackage[libertine]{newtxmath}
\usepackage{color}
\usepackage{dsfont, hyperref, xcolor}
\usepackage{comment}
\usepackage{enumerate}
\newcommand{\Tr}[1]{\text{Tr}\left\{#1\right\}}

\begin{document}

\title{Normal and superconducting currents through
the Sachdev-Ye-Kitaev model}

\author{Gianluca~Francica, Marco Uguccioni, Luca Dell'Anna}
\address{Dipartimento di Fisica e Astronomia e Sezione INFN, Universit\`{a} degli Studi di Padova, via Marzolo 8, 35131 Padova, Italy}

\date{\today}

\begin{abstract}
We study the current driven by an applied voltage as a function of time through 
the Sachdev-Ye-Kitaev model 
when coupled to two normal or superconducting reservoirs. 
For normal leads, in the strong coupling limit and for small bias, the current through the Sachdev-Ye-Kitaev model, described by a quartic interaction term, reaches monotonically the stationarity, in contrast to the case of a disordered quadratic interaction where the current has a peak before reaching the stationary phase. For superconducting leads the currents have oscillations whose frequencies are determined by the gap and the voltage, and are suppressed in the strong coupling limit. Moreover, due to different short time scales between the normal and the oscillating part of the superconducting current, a peak appears before reaching the stationarity.
\end{abstract}

\maketitle

\section{Introduction}

The Sachdev-Ye-Kitaev (SYK) model has attracted a lot of interest in recent years \cite{SachdevYe, kitaevtalk}. It displays very different features with respect to common Fermi liquids (see, e.g., Ref.~\cite{Chowdhury22} for a review), for instance, the resistivity is linear with respect to the temperature, exhibiting a so-called Planckian transport~\cite{Chowdhury22,Patel19,Sachdev23}. Moreover, the SYK model is dual to a two-dimensional nearly anti-de Sitter space~\cite{kitaevtalk,kitaev18,maldacena16,sarosi19}, opening a different way to investigate black holes.
Concerning its experimental realization, there are already many proposals either in solid state physics (see, e.g., Ref.~\cite{Franz18} for a review) and in cavity quantum electrodynamics platforms \cite{hauke}.
Although several studies have been done investigating the mesoscopic physics by the SYK model
(see, e.g., Refs.~\cite{song17,davison,Gnezdilov18,Can19,Guo20,Kulkarni22}), a characterization of the current and supercurrent, driven by a double contact setup, as functions of time is still lacking, differently from the case of a single colder bath, where it is known that the SYK heats up at short times, before thermalizing~\cite{Almheiri19,Zhang19,Cheipesh21}. Concerning its isolated dynamics, also the case of a quantum quench has been investigated~\cite{Eberlein17}. {\color{black} Other recent studies concern eternal traversable wormholes~\cite{Zhou20}, the Bekenstein-Hawking entropy~\cite{Kruchkov20} and  the existence of anomalous power laws in the temperature dependent conductance~\cite{Altland19}.}

We calculate the current triggered by out-of-equilibrium normal and superconducting leads, and show some peculiarities 
never observed so far. 
We consider a time-dependent tunneling where the voltage enters via a time-dependent phase, e.g., see Ref.~\cite{Jacquet20}.
For a weak tunneling coupling, we can get metastability, signaled by a peak at short times, due to a separation of timescales (see, e.g., Ref.~\cite{Macieszczak21}), which happens for normal leads in a disordered Fermi liquid when the disorder is large enough. In contrast, for the SYK, the current monotonically increases without showing metastability for any value of the disorder strength.
Furthermore, for superconducting leads we get oscillations in time with frequencies that can be affected also by the timescale of the medium. For the SYK model these frequencies are determined by the gap and the applied voltage.

The paper is organized as follows: in Sec. II we briefly introduce the SYK model, in Sec. III we first calculate in general terms the expression of the current, with anomalous components for the superconducting leads, and then present its form in the weak coupling limit, used to derive the results reported in Sec. IV. The final section is devoted to conclusions.
\section{SYK Model}
We consider $N$ Majorana fermions $\chi_j$, satisfying the anticommutation relations $\{\chi_j,\chi_k\}=\delta_{jk}$, with disordered all-to-all $q$-body interactions described by the SYK Hamiltonian, which, for $q=4$ fields (let us denote it SYK$_4$), reads
\begin{equation}
H_0=-\frac{1}{4!}\sum_{j,k,l,m} J_{jklm} \chi_j \chi_k \chi_l\chi_m
\end{equation}
where $J_{jklm}$ have a Gaussian distribution such that $\overline{J^2_{jklm}} = 3! J^2/N^3$.
At a single site $j$ the Keldysh Green function averaged over the disorder reads
\begin{equation}
G_0(t,t')= - \langle T \chi (t) \chi (t')\rangle
\end{equation}
where $\chi=\chi_j$ and $T$ is the time ordering operator with respect to the contour $C'=i[\beta,0^+)\cup C$ where $C=[i 0^+,\tau+i 0^+)\cup (\tau+i 0^-,i 0^-]$. By using the dynamical mean field theory as $N\to \infty$ we get the equations
\begin{eqnarray}
\label{eq.SD1}G_0(t,t') &=& (-\partial_t+\Sigma_0(t,t'))^{-1}\\
\label{eq.SD2}\Sigma_0(t,t') &=& J^2 G_0^3(t,t')
\end{eqnarray}
which can be solved numerically, as explained in Appendix~\ref{app.numsol}.

\section{Tunneling}
We consider a tunneling with two superconductive leads ($L$ and $R$) switched on at times $t>0$, described by the Hamiltonian $H_T = H_{T,L}+H_{T,R}$ where
\begin{equation}
H_{T,a} =  \sum_{k,j} w_{akj}e^{is_a Vt/2} (c^\dagger_{ak\uparrow}+ c^\dagger_{ak\downarrow})\chi_j + H.c.
\end{equation}
where $a=L,R$ and $s_{L/R}=\pm$. In detail, $c^\dagger_{ak\sigma}$ are the creation operators of the leads $a=L,R$ and $w_{akj}$ are the tunneling amplitudes which are random with a Gaussian distribution having variance $\sigma_w^2\sim 1/N$. The bias voltage $V$ enters via a time-dependent phase.
The two leads  are d dimensional s-wave superconductors described by the Hamiltonian
\begin{equation}
H_a = \sum_k \Psi^\dagger_{ak} \left(
                              \begin{array}{cc}
                                \xi_k & -\Delta \\
                                -\Delta & -\xi_k \\
                              \end{array}
                            \right) \Psi_{ak}
\end{equation}
where we defined the Nambu spinor $\Psi_{ak} = (c_{a k \uparrow},c_{a -k \downarrow})^T$. 
At the initial time the total system, with Hamiltonian $H_{tot}(0)=H_0+H_L+H_R$, is in the equilibrium state at the inverse temperature $\beta$ with density matrix $\rho_{tot}=e^{-\beta H_{tot}}/Z_{tot}$, where $Z_{tot}=\Tr{e^{-\beta H_{tot}}}$ is the partition function.
The Hamiltonian of the leads can be diagonalized by performing the Bogoliubov transformation
\begin{equation}
\left(
  \begin{array}{c}
    \alpha_{ak\uparrow} \\
    \alpha^\dagger_{a-k\downarrow} \\
  \end{array}
\right) = \left(
            \begin{array}{cc}
              \cos \theta_k & \sin \theta_k \\
              \sin \theta_k & -\cos \theta_k \\
            \end{array}
          \right) \left(
                    \begin{array}{c}
                      c_{a k \uparrow} \\
                      c^\dagger_{a -k \downarrow} \\
                    \end{array}
                  \right)
\end{equation}
with $\cos(2\theta_k) = \xi_k/\lambda_k$, $\sin(2\theta_k) = -\Delta/\lambda_k$, and $\lambda_k = \textrm{sign}(\xi_k)\sqrt{\xi_k^2+\Delta^2}$, such that
\begin{equation}
H_a= \sum_{k,\sigma}\lambda_k \alpha_{a k\sigma}^\dagger\alpha_{ a k\sigma}
\end{equation}
The left current is $J_{L}(t)= J_{L,\uparrow}(t) + J_{L,\downarrow}(t)$ where
\begin{equation}
J_{L,\sigma}(t) = i \sum_{k,j} w_{Lkj}  e^{iVt/2} \langle c_{Lk\sigma}^\dagger \chi_j (t) \rangle - e^{-iVt/2} \langle \chi_j  c_{Lk\sigma}(t) \rangle
\end{equation}
For the spin up current we get the Jauho, Wingreen, and Meir formula~\cite{Jauho94}
\begin{eqnarray}
\nonumber J_{L\uparrow}(t) = 2  \text{Re}\sum_{k,j} w_{Lkj} e^{iVt/2} (\cos \theta_k G^<_{L k \uparrow j}(t,t) \\
+ \sin \theta_k F^<_{L-k \downarrow j}(t,t))
\end{eqnarray}
where
\begin{eqnarray}
G^<_{L k \sigma j}(t,t')=i\langle \alpha^\dagger_{Lk\sigma}(t') \chi_j(t)\rangle\\
F^<_{L k \sigma j}(t,t')=i\langle \alpha_{Lk\sigma}(t') \chi_j(t)\rangle
\end{eqnarray}
A similar equation is achieved for the spin down current,
\begin{eqnarray}
\nonumber J_{L\downarrow}(t) = 2 \text{Re}\sum_{k,j} w_{Lkj} e^{iVt/2} (-\cos \theta_k G^<_{L -k \downarrow j}(t,t) \\
+ \sin \theta_k F^<_{Lk \uparrow j}(t,t))
\end{eqnarray}
Since the leads are noninteracting, we get
\begin{eqnarray}
\nonumber G^<_{L k \uparrow j}(t,t')&=&i \sum_{j'} w_{Lkj'} \int_C dt_1 g_{Lk\uparrow}(t_1,t') G_{jj'}(t,t_1)\\
&& \times \, (e^{-i Vt_1/2}\cos \theta_k-e^{i Vt_1/2}\sin \theta_k) \\
\nonumber \\
\nonumber F^<_{L k \downarrow j}(t,t')&=&-i \sum_{j'} w_{Lkj'} \int_C dt_1 g_{L k\downarrow}(t',t_1) G_{jj'}(t,t_1)\\
&& \times \, (e^{i Vt_1/2}\cos \theta_k+e^{-i Vt_1/2}\sin \theta_k) \\
\nonumber \\
\nonumber G^<_{L k \downarrow j}(t,t')&=&-i \sum_{j'} w_{Lkj'} \int_C dt_1 g_{Lk\downarrow}(t_1,t') G_{jj'}(t,t_1)\\
&&\times \, (e^{-i Vt_1/2}\cos \theta_k+e^{i Vt_1/2}\sin \theta_k) \\
\nonumber \\
\nonumber F^<_{L k \uparrow j}(t,t')&=&i \sum_{j'} w_{Lkj'} \int_C dt_1 g_{L k\uparrow}(t',t_1) G_{jj'}(t,t_1)\\
&&\times \, (e^{i Vt_1/2}\cos \theta_k-e^{-i Vt_1/2}\sin \theta_k)
\end{eqnarray}
where we have defined the bare Keldysh Green functions of the leads
\begin{equation}
g_{Lk\sigma}(t,t') = -i \langle T \alpha_{Lk\sigma}(t) \alpha^\dagger_{Lk\sigma}(t')\rangle
\end{equation}
which can be evaluated as usual
\begin{eqnarray}
\label{eq. g1} && \hspace{-0.5cm}
g^<_{ak\sigma}(t,t') = i  \langle \alpha_{ak\sigma}^\dagger(t') \alpha_{ak\sigma}(t)\rangle = i n_{ak} e^{-i\lambda_k(t-t')}\\
\label{eq. g2} && \hspace{-0.5cm}
g^>_{ak\sigma}(t,t') = i (n_{ak}-1)e^{-i\lambda_k(t-t')}\\
\label{eq. g3} && \hspace{-0.5cm}
g^T_{ak\sigma}(t,t') = \theta(t-t') g^>_{ak\sigma}(t,t') + \theta(t'-t) g^<_{ak\sigma}(t,t')\\
\label{eq. g4} && \hspace{-0.5cm}
g^{\tilde T}_{ak\sigma}(t,t') = \theta(t'-t) g^>_{ak\sigma}(t,t') + \theta(t-t') g^<_{ak\sigma}(t,t')
\end{eqnarray}
where $n_{L/R k} = 1/(1+e^{\beta \lambda_k})$. The Keldysh Green function of the central region is defined as
\begin{equation}
G_{jj'}(t,t')= - \langle T \chi_j (t) \chi_{j'} (t')\rangle
\end{equation}
without averaging over the disorder.
Thus, we get
\begin{eqnarray}\label{eq. JL cos 11 0}
\nonumber J_{L}(t) = 2\text{Re} \sum_{k,j,j'} w_{Lkj}w_{Lkj'} e^{iVt/2}i\int_Cdt_1e^{-iVt_1/2} G_{jj'}(t,t_1) \\
 \times \big((1+\cos 2\theta_k) g_{Lk\sigma}(t_1,t)-(1-\cos 2\theta_k)g_{Lk\sigma}(t,t_1)\big)\;
\end{eqnarray}
where the argument $t$ of $G_{jj'}(t,t_1)$ belongs to the forward branch $[i 0^+,\tau+i 0^+)$ and the argument $t$ of $g_{Lk\sigma}$ belongs to the backward branch $(\tau+i 0^-,i 0^-]$.
{\color{black} We aim to perform the average over disorder of the current in Eq.~\eqref{eq. JL cos 11 0}. Thus, we separate the sum over $j$ and $j'$ in two sums, i.e., $\sum_{j,j'}=\sum_{j=j'}+\sum_{j\neq j'}$, achieving $J_{L}(t) =  J_{L,d}(t) + J_{L,od}(t)$, where in the term $J_{L,d}(t)$ we sum over the indices $j=j'$, while in $J_{L,od}(t)$ we sum over the remaining indices $j\neq j'$.
As explained in Appendices~\ref{app.action} and \ref{app.avecur}, we can perform the average over all the disorder getting
\begin{equation}
\overline{J}_{L}(t) =  \overline{J}_{L,d}(t) + \overline{J}_{L,od}(t)\,,
\end{equation}
where the first term is
\begin{eqnarray}
\nonumber \overline{J}_{L,d}(t) = 2N\sigma_w^2 \,\text{Re} \sum_k e^{iVt/2}i
\int_Cdt_1e^{-iVt_1/2} G(t,t_1)\\
\label{eq. JL cos}\times \big((1+\cos 2\theta_k)g_{Lk\sigma}(t_1,t)-(1-\cos 2\theta_k)g_{Lk\sigma}(t,t_1)\big)\,,
\end{eqnarray}
with $G(t,t_1)$ the disorder average of $G_{jj}(t,t_1)$, 
calculated by using the effective action in Eq.~\eqref{eq. S eff}, 
which is equal to $G_0$ in the thermodynamic limit 
(see Appendix~\ref{app.avecur}). 
The second term is, instead, 
\begin{eqnarray}
\nonumber \overline{J}_{L,od}(t) = 2 N^2 \sigma_w^4\,\text{Re} \sum_{k}  e^{iVt/2}i\int_Cdt_1e^{-iVt_1/2} \,\widetilde{G}(t,t_1)\\
\nonumber \times \big((1+\cos 2\theta_k) g_{Lk\sigma}(t_1,t)-(1-\cos 2\theta_k)g_{Lk\sigma}(t,t_1)\big) \,,
\label{eq. JL cos 11 off-d}
\end{eqnarray}}
where $\widetilde{G}(t,t_1)$ is defined in Eq.~(\ref{eq.calcolo}). 

Within a constant density of states approximation, $\rho(\xi)\simeq \rho$,
Eq.~\eqref{eq. JL cos} simplifies as follows
(see Appendix \ref{app.green})
\begin{equation}\label{eq.JLfinal}
\overline{J}_{L,d}(t) = 4N\sigma_w^2 \text{Re} \sum_k e^{iVt/2}i\int_Cdt_1e^{-iVt_1/2}g_{Lk\sigma}(t_1,t)G(t,t_1)
\end{equation}
so that we need to calculate only terms like
\begin{equation}\label{eq. intless}
\sum_k g^<_{Lk\uparrow}(t_1,t) = i \rho \left(\frac{L}{\pi}\right)^d \int_I d\lambda \frac{|\lambda|}{\sqrt{\lambda^2-\Delta^2}} \frac{e^{-i\lambda(t_1-t+i 0^+)}}{1+e^{\beta\lambda}}
\end{equation}
and
\begin{equation}\label{eq. intgreat}
\sum_k g^>_{Lk\uparrow}(t_1,t) = -i \rho \left(\frac{L}{\pi}\right)^d \int_I d\lambda \frac{|\lambda|}{\sqrt{\lambda^2-\Delta^2}} \frac{e^{-i\lambda(t_1-t-i 0^+)}}{1+e^{-\beta\lambda}}
\end{equation}
where $I=(-\infty,-\Delta]\cup [\Delta,\infty)$.
For $\Delta$ small, we get
\begin{equation}
\sum_k g^<_{Lk\uparrow}(t_1,t) = i \rho \left(\frac{L}{\pi}\right)^d \int_I d\lambda \frac{e^{-i\lambda(t_1-t+i 0^+)}}{1+e^{\beta\lambda}}
\end{equation}
which can be evaluated exactly,
\begin{equation}\label{eq. inte 1}
\beta\int_I d\lambda \frac{e^{-i\lambda(t_1-t+i 0^+)}}{1+e^{\beta \lambda}}= I^<_1+I_2^<+\frac{2i\pi}{\sinh(\pi(t_1-t)/\beta)}
\end{equation}
where
\begin{eqnarray}
I^<_1 &=& e^{\pi (t_1-t)/\beta }B_z(1+i(t_1-t)/\beta,0)\\
I^<_2 &=& -e^{-\pi (t_1-t)/\beta }B_z(-i(t_1-t)/\beta,0)
\end{eqnarray}
with $z=-e^{\beta\Delta}$ and $B_z$ is the incomplete beta function, 
$B_z(a,b) = \int_0^z dt\, t^{a-1}(1-t^{b-1})$.
Analogously, for $\Delta$ small, Eq.~(\ref{eq. intgreat}) becomes 
\begin{equation}
\sum_k g^>_{Lk\uparrow}(t_1,t) = -i \rho \left(\frac{L}{\pi}\right)^d \int_I d\lambda \frac{e^{-i\lambda(t_1-t-i 0^+)}}{1+e^{-\beta\lambda}}\,.
\end{equation}
The integral can be performed exactly and we get
\begin{equation}\label{eq. inte 2}
\beta\int_I d\lambda \frac{e^{-i\lambda(t_1-t-i 0^+)}}{1+e^{-\beta \lambda}}= I^>_1+I^>_2-\frac{2 i \pi}{\sinh(\pi(t_1-t)/\beta)}
\end{equation}
where
\begin{eqnarray}
I^>_1 &=& e^{-\pi (t_1-t)/\beta }B_z(1-i(t_1-t)/\beta,0)\\
I^>_2 &=& \frac{i\beta e^{i\Delta (t_1-t)}}{(t_1-t)}{}_2 F_1\big(1,i(t_1-t)/\beta;1+i(t_1-t)/\beta,z\big)
\end{eqnarray}
where ${}_2 F_1$ is the 
hypergeometric function.
For the right current $\overline{J}_R$ it is enough to replace $V$ with $-V$, so that $\overline{J}_R(t)=-\overline{J}_L(t)$.
For simplicity we define the current density
\[j\equiv \pi^{d-1}\overline{J}_L/(N\sigma_w^2 \rho L^d)\]
which is a dimensionless quantity.
In what follows we will focus on a weak tunneling limit, $N\sigma_w^2\ll 1$. The leading term in the current $\overline{J}_L$ is, then, {\color{black}$\overline{J}_{L,d}$, namely 
\begin{equation}
\overline{J}_L \simeq \overline{J}_{L,d}, 
\end{equation}
while $\overline{J}_{L,od}$ dominates for strong tunneling.}

\section{Results}
Let us calculate the current as a function of time after switching on the tunneling and the bias voltage, considering normal and superconducting leads.
\subsection{Normal leads}
\begin{figure}
[t!]
\includegraphics[width=0.499\columnwidth]{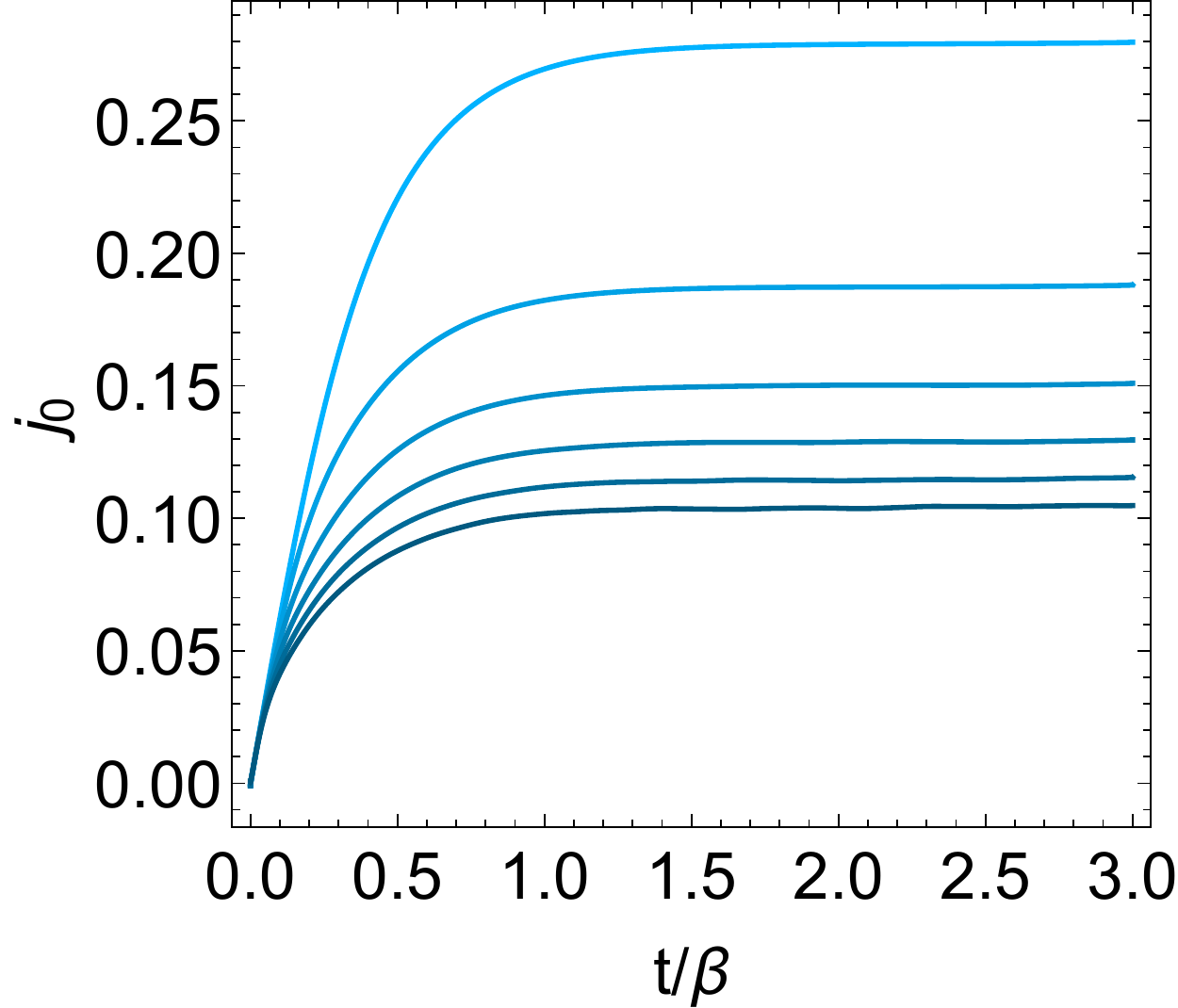}\includegraphics[width=0.499\columnwidth]{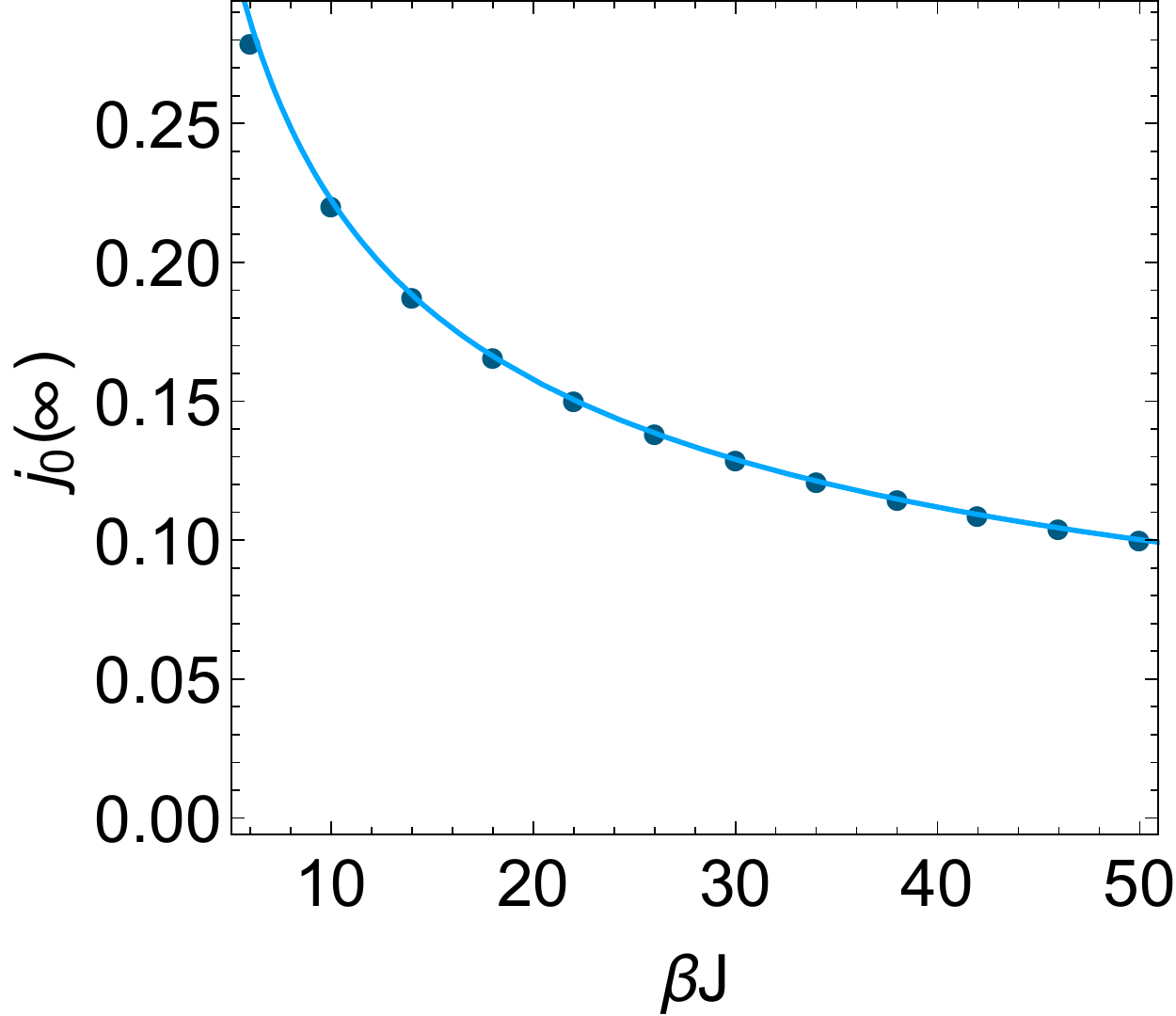}\\
\includegraphics[width=0.499\columnwidth]{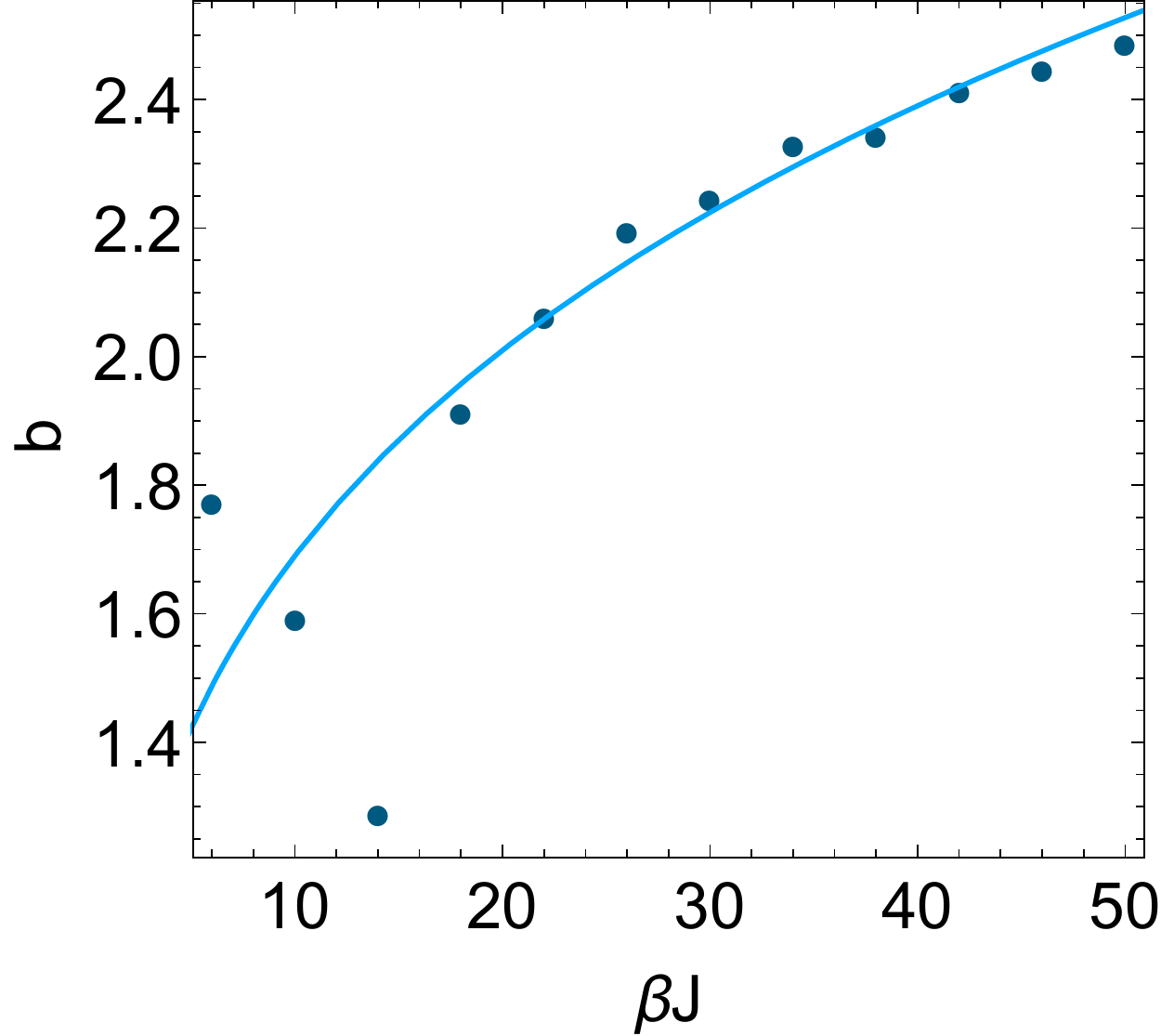}\includegraphics[width=0.499\columnwidth]{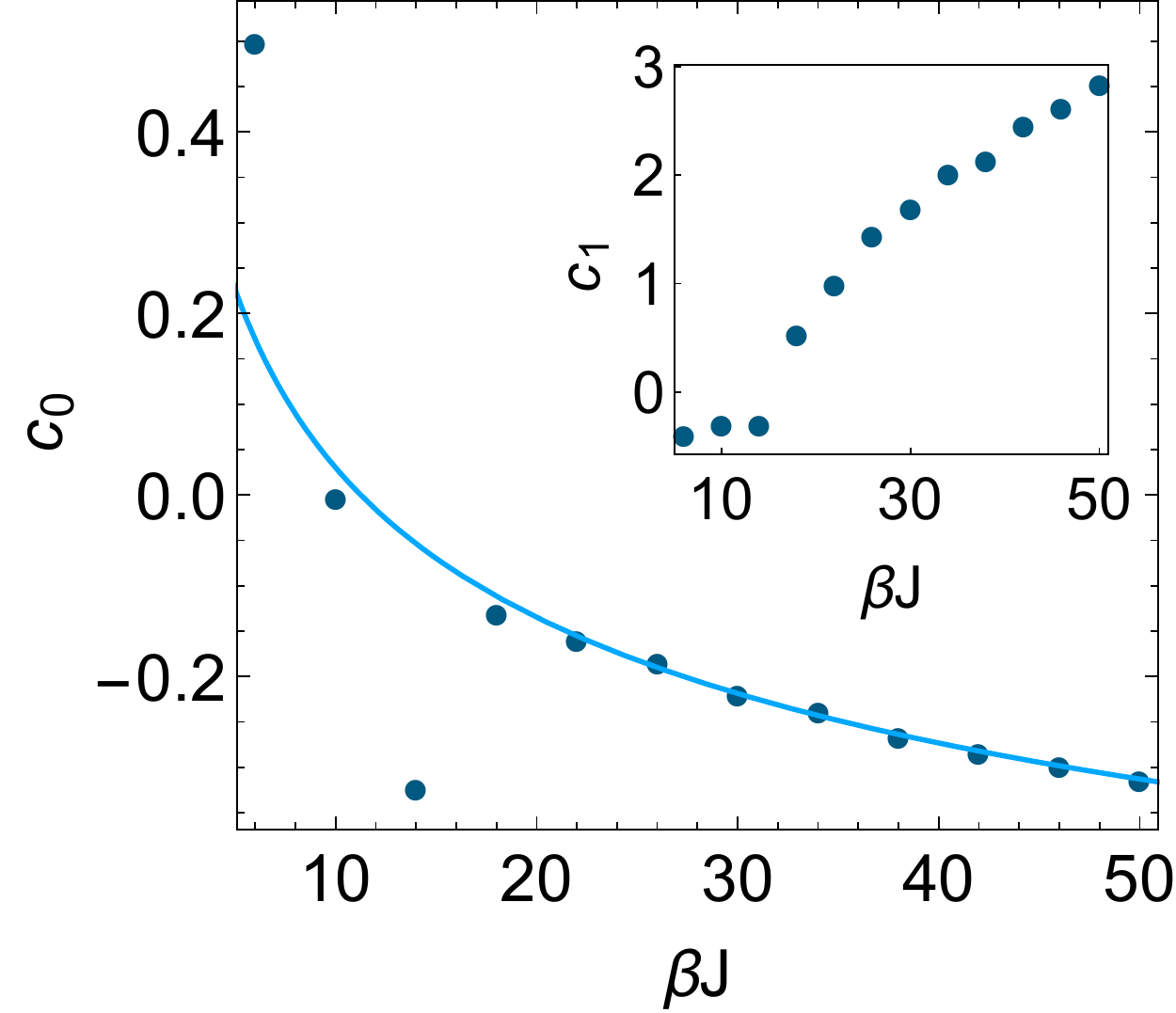}
\caption{{\color{black} SYK$_4$ with normal leads:} (Top left panel) The current $j_0$ as a function of time for different values of $\beta J$ and small voltage. We put $\beta V=1$ and $\beta J=6,14,22,30,38,46$ (from light to dark lines). The best fit with respect to the function in Eq.~\eqref{eq. tanh modi} is indistinguishable by eye from the data-points. 
(Top right panel) The best fit for $j_0(\infty)$ using the form $u J^z$. We get $u\approx 0.7$ and $z\approx -0.5$ ($\beta=1$).
(Bottom left panel) For $b$ we find the best fit with $u J^{1/4}$, getting $u\approx1$.
(Bottom right panel) For $c_0$ we find the best fit with $u J^x-1$, getting $x\approx -0.25$, and $u\approx 1.8$. In the inset, $c_1$ as a function of $J$ is also reported. For all the fits we consider the points $\beta J>20$.  }
\label{fig:plot1}
\end{figure}

We start focusing on small voltage $\beta V \lesssim 1$. For normal leads, namely for $\Delta=0$, the current is plotted in Fig.~\ref{fig:plot1}.
We find that for a relatively small coupling, $\beta J \lesssim 1$, the current $j_0(t)\equiv j(t)$ (to distinguish from $\Delta\neq 0$ case) behaves as
\begin{equation}\label{eq.tanh}
j_0(t)\simeq  a \tanh(b t/\beta)\,.
\end{equation}
Upon increasing $J$,
the current, as a function of the time, is modified and is better approximated by
\begin{equation}\label{eq. tanh modi}
j_0(t) \simeq a \frac{e^{b t/\beta}- e^{-b t/\beta}}{e^{b t/\beta}+ (c_0+c_1 t/\beta) e^{-b t/\beta}}
\end{equation}
The parameters $a$, $b$, $c_0$ and $c_1$ determined by finding the best fit are plotted as functions of $J$ in Fig.~\ref{fig:plot1}. We see that for large coupling the stationary current goes as $a\sim 1/\sqrt{J}$ in agreement with the conformal solution. Furthermore, we find $c_0\sim u J^{-1/4} -1$ and $b \sim J^{1/4}$.
For very small times $t/\beta \ll 1$, from Fig.~\ref{fig:plot1} we see that $j(t)$ is linear with the same slope for any $J$, i.e., $j_0 \sim  \alpha t/\beta$, where $\alpha$ does not depend on $J$. In particular, from a linear fit we get $\alpha \approx \pi \beta V /5$ for small $V$. Actually from Eq.~\eqref{eq. tanh modi}, we get
\begin{equation}
\alpha \simeq \frac{2ab}{1+c_0}
\end{equation}
so that by using the asymptotic formulas of the parameters for large $J$, we get $\alpha$ constant in $J$. 

\begin{figure}
[t!]
\includegraphics[width=0.499\columnwidth]{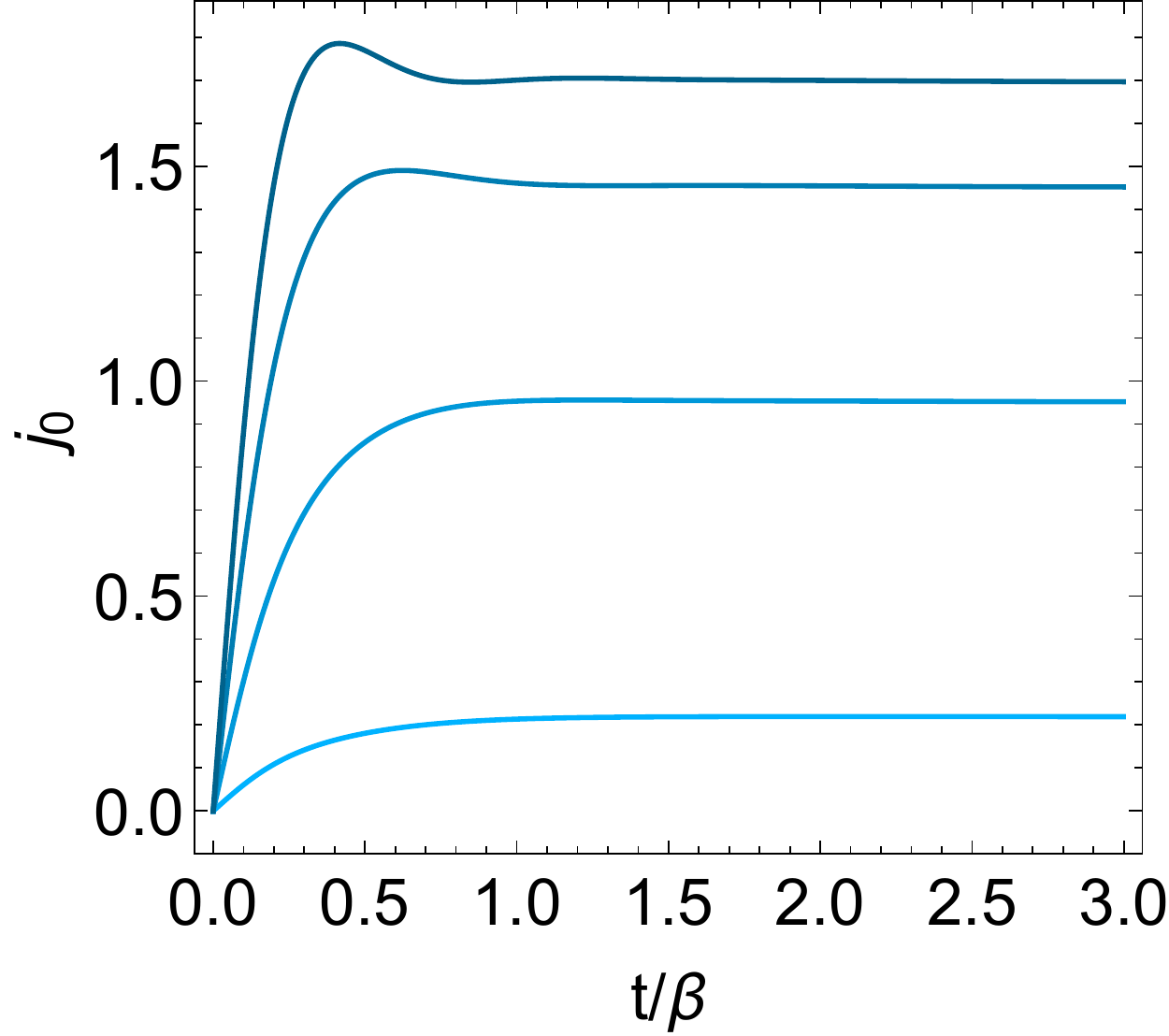}\includegraphics[width=0.499\columnwidth]{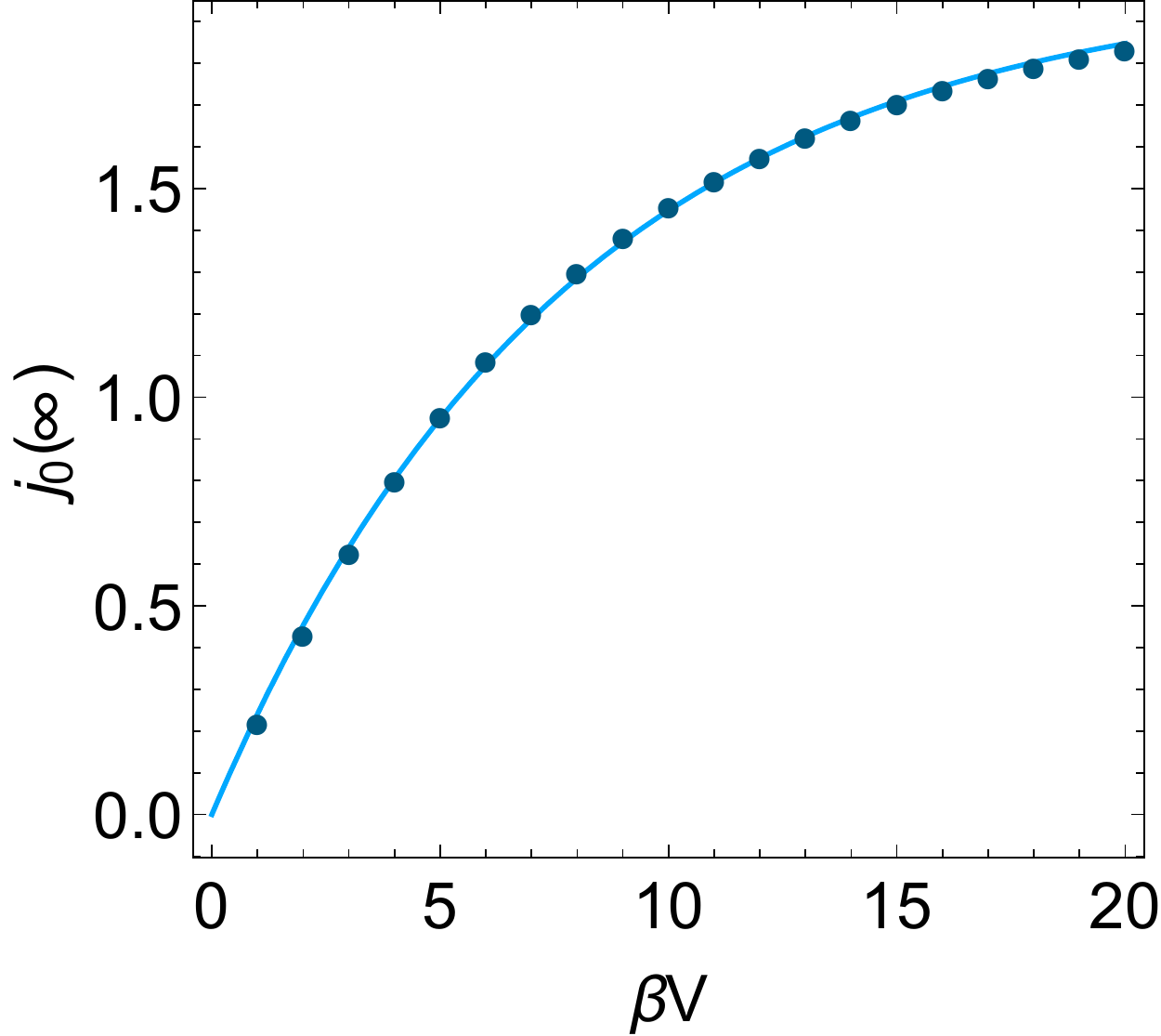}
\caption{{\color{black} SYK$_4$ with normal leads:} (Top left panel) The current $j_0$ as a function of time for different values of $\beta V$. We put $\beta J=10$ and $\beta V=1,5,10,15$ (from light to dark lines). (Top right panel) The stationary values of the current, calculated as the temporal average for times $5<t/\beta<10$, for different values of $V$. The solid line is obtained by finding the best fit using the function in Eq.~\eqref{eq. j vs V}.}
\label{fig:plot2}
\end{figure}

Differently from the case of small voltages where the current monotonically increases in time, for very large $V$ the current displays a peak at short times (see Fig.~\ref{fig:plot2}), because a separation of time scales occurs.
Furthermore, the stationary current as a function of the voltage saturates to $2$,
independently on $J$, by inspection of Eq.~(\ref{eq.JLfinal}),
since, for large $V$,  the term $e^{iV(t-t_1)/2}$ 
oscillates rapidly, and we can perform the approximation $G_0(t,t')\approx -1/2$ and calculating the integral over $t_1$, for long $t$ and large $V$ we get $j(\infty)=2$.
In particular, as shown in Fig.~\ref{fig:plot2} the stationary current goes like
\begin{equation}\label{eq. j vs V}
j(\infty)\simeq 2(1-e^{-c\beta V})\,.
\end{equation}
We find that the differential conductance goes exponentially to zero as the bias increases.
For a small bias, $j(\infty) \sim c \beta V$, and looking at Fig.~\ref{fig:plot1} (top right panel), in the same regime of small voltage, 
we have $c \sim 1/\sqrt{\beta J}$.
Moreover, for a small bias, we get $j(\infty)\sim  V \sqrt{\beta/J}$, therefore, the resistivity goes as $\sim \sqrt{J/\beta}$, i.e., as  $\sqrt{T}$, as square root of the temperature, in perfect agreement with what observed in Ref.~\cite{Can19}.

Finally we notice that, for $q=2$, i.e. for a SYK$_2$ model, by similar calculations, from the two-point correlation function,
we get $j(\infty)\sim \beta V ((\beta J)^{-1} + c (\beta J)^{-3}) $, thus a resistivity $\sim \rho_0 + c \beta^{-2}$ which is quadratic in the temperature, in agreement with the conventional Fermi liquid result. 

\subsection{Superconducting leads}
\begin{figure}
[t!]
\includegraphics[width=0.499\columnwidth]{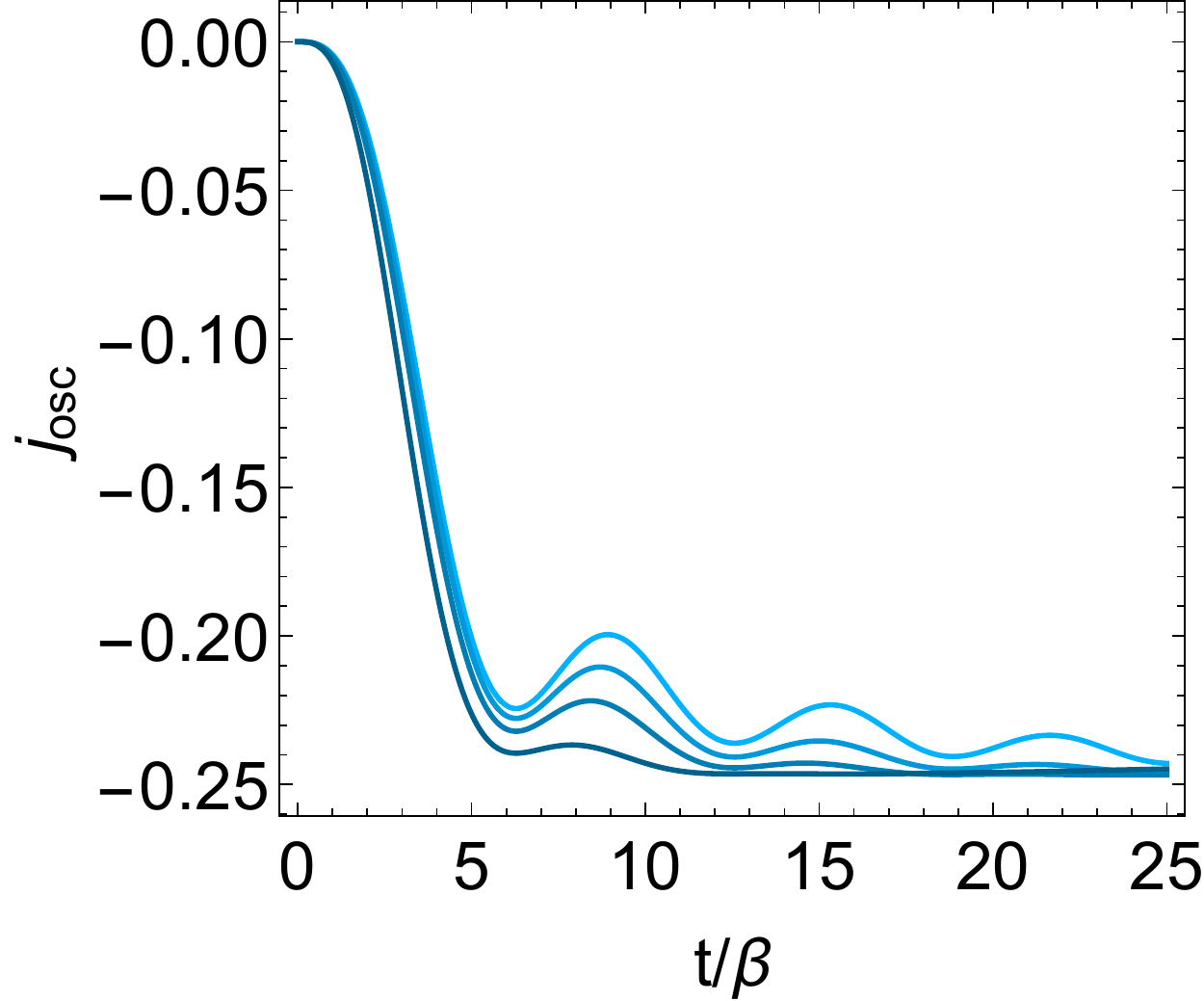}\includegraphics[width=0.499\columnwidth]{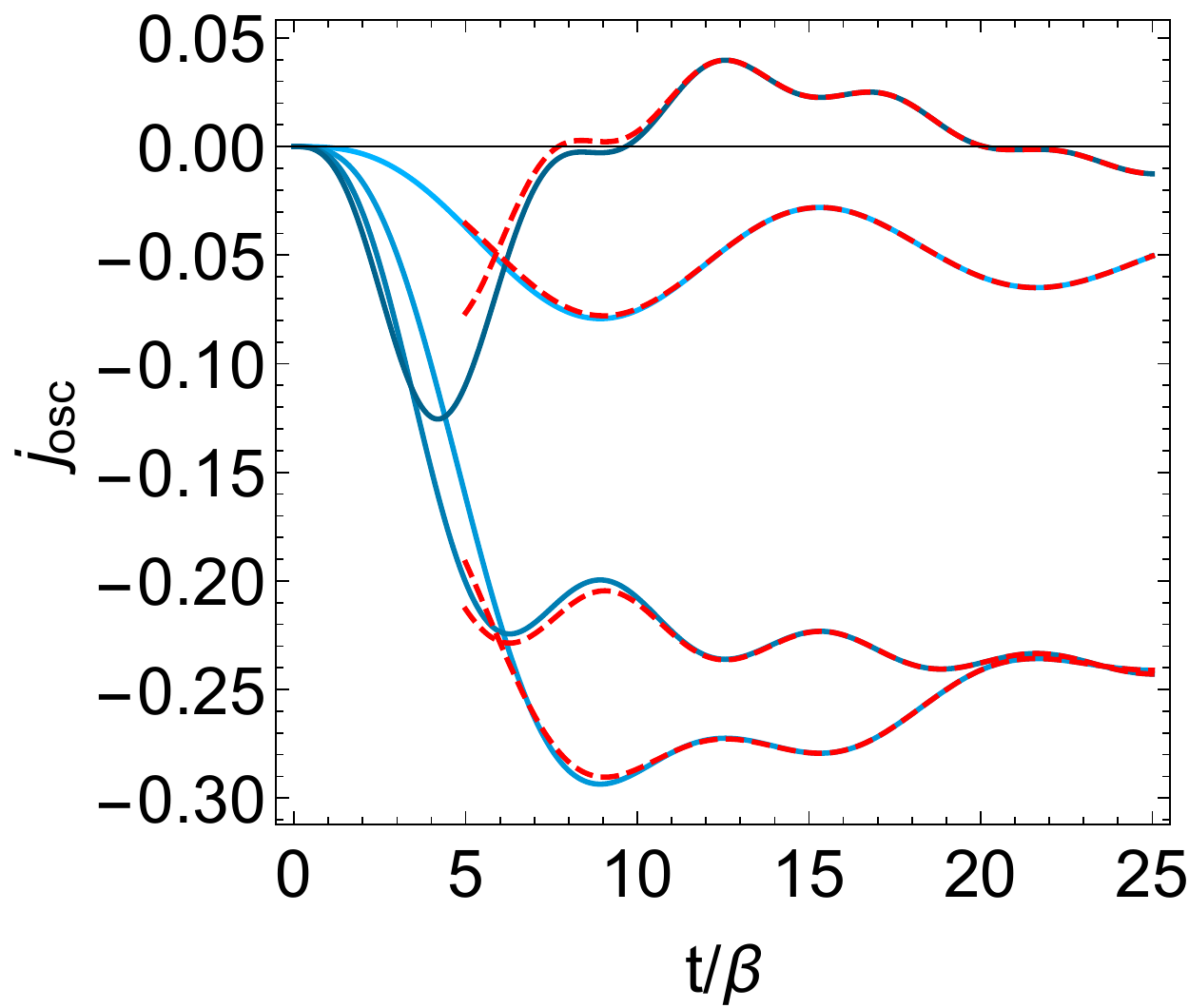}
\caption{{\color{black} SYK$_4$ with superconductive leads:} (Left panel) The current contribution $j_{osc}$ as a function of time for different values of $\beta J$. We put $\beta V=1$, $\beta \Delta=0.5$ and $\beta J=0.1,0.2,0.3,0.5$ (from light to dark lines). (Right panel) $j_{osc}$ as a function of time for different values of $\beta V$. We put $\beta J=0.1$, $\beta \Delta=0.5$ and $\beta V=0.1,0.5,1.,1.5$ (from light to dark lines). The red dashed lines are obtained by finding the best fit using the function in Eq.~\eqref{eq. j oscillations}, for $15<t/\beta<25$.} 
\label{fig:plot2.5}
\end{figure}

\begin{figure}
[t!]
\includegraphics[width=0.499\columnwidth]{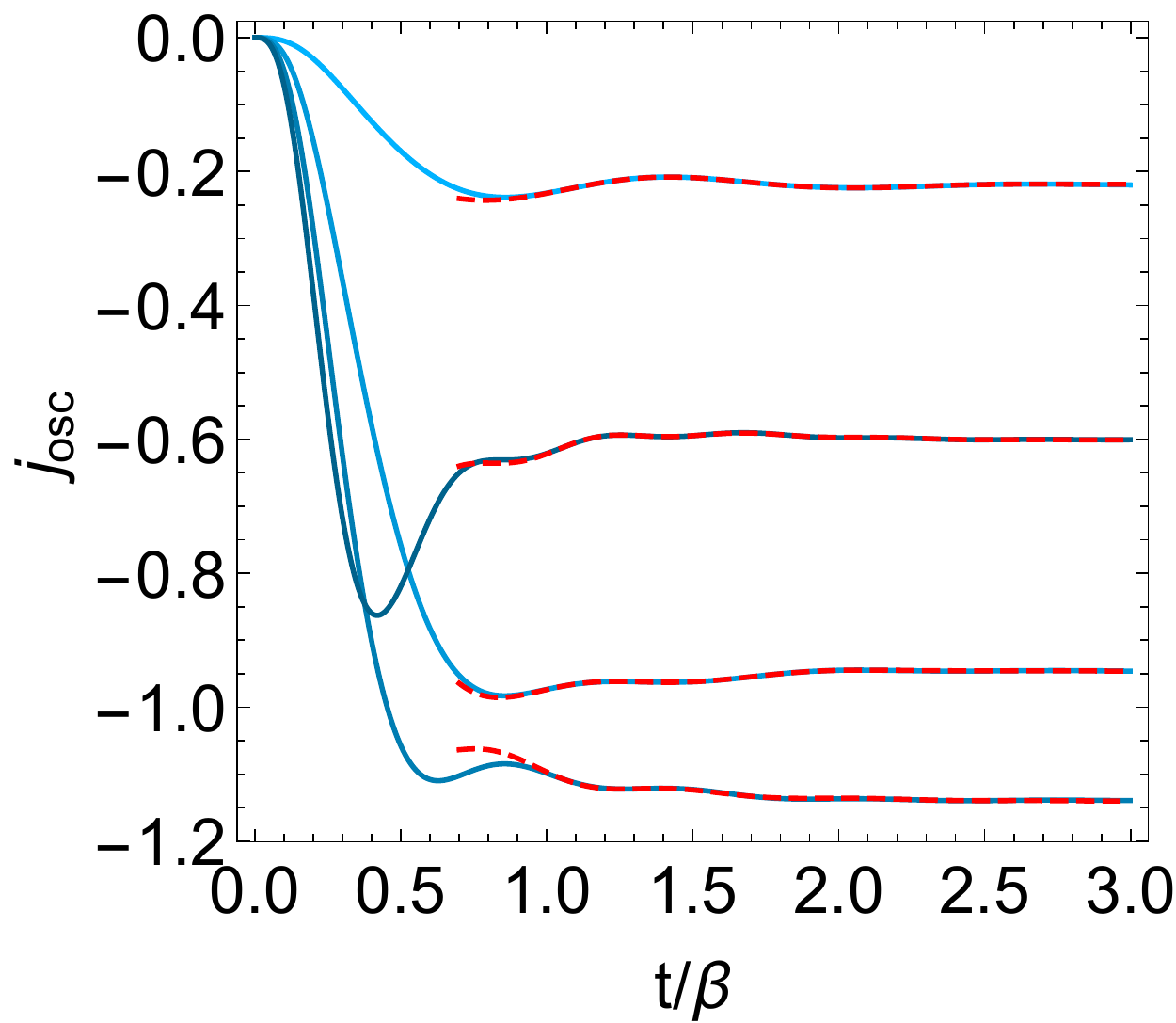}\includegraphics[width=0.499\columnwidth]{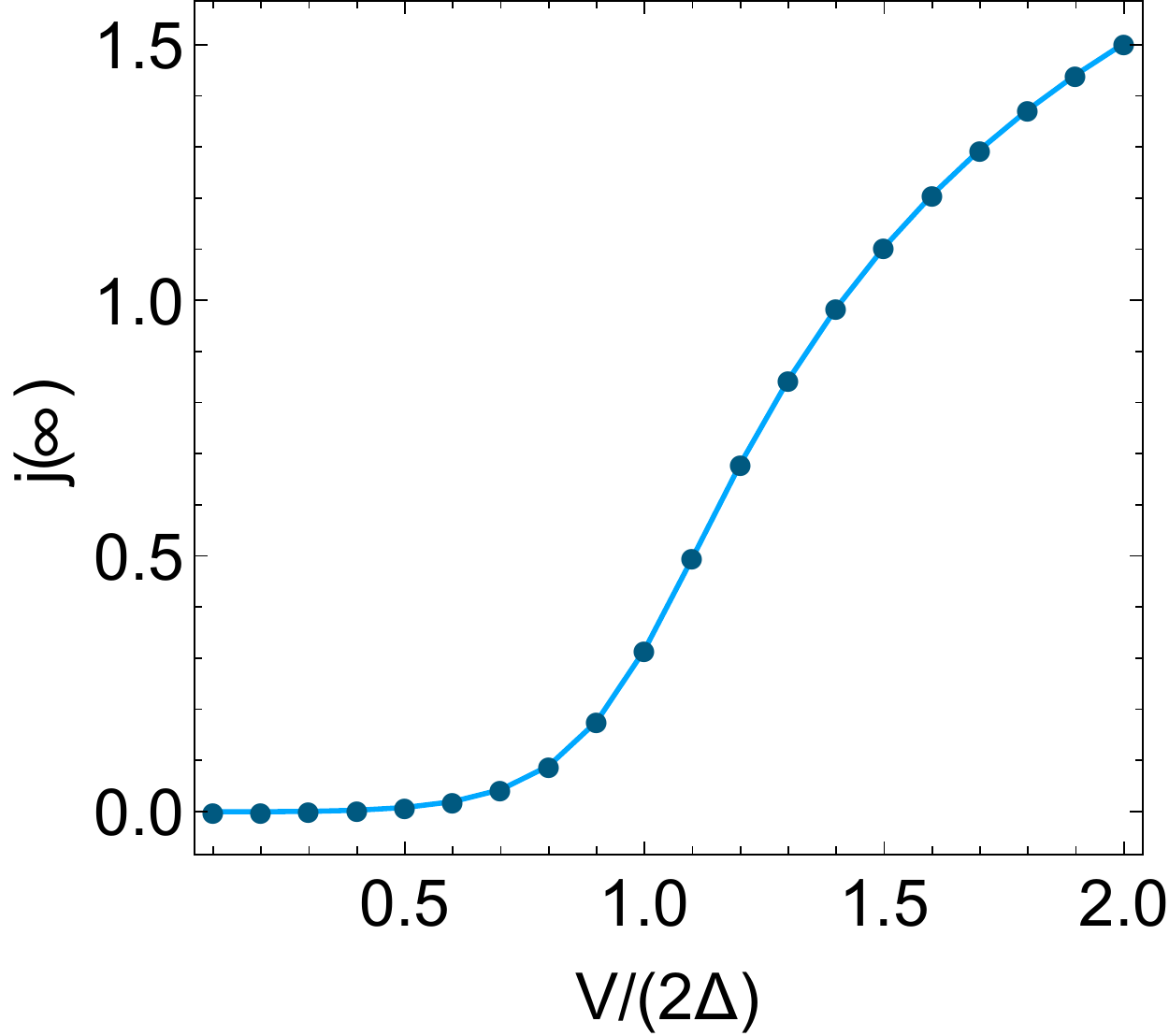}
\caption{{\color{black} SYK$_4$ with superconductive leads:} (Left panel) the current $j_{osc}$ as a function of time for different values of $\beta V$. We put $\beta J=10$, $\beta \Delta=5$ and $\beta V=1,5,10,15$ (from light to dark lines). The red dashed line are obtained by finding the best fit using the function in Eq.~\eqref{eq. j oscillations}
for $1<t/\beta<3$. 
(Right panel) the stationary value of the full current, $j(\infty)$, calculated averaging over time, for $5<t/\beta<10$, as a function of $V$ in units of $2\Delta$.}
\label{fig:plot3}
\end{figure}

Let us consider the SYK model contacted with superconducting leads. In general, for $\Delta\neq 0$, we can isolate the terms coming from $1/\sinh(\pi(t_1-t)/\beta)$, i.e., the last terms in Eqs.~\eqref{eq. inte 1} and~\eqref{eq. inte 2}, which give the contribution to the current equal to $2j_0(t)$ which does not depend on $\Delta$. We can write
\begin{equation}
j(t) = 2j_0(t) + j_{1}(t)+ j_{2}(t)
\end{equation}
where $j_1$ comes from $I^<_1$ and $I^>_1$, and it is equal to zero when $\Delta=0$, while $j_{2}(t)$ comes from $I^<_2$ and $I^>_2$ and it is equal to $-j_0$ when $\Delta =0$. Thus, we can rewrite
\begin{equation}
j(t) = j_0(t) + j_{osc}(t)
\end{equation}
where
\begin{equation}
 j_{osc}(t)= j_0(t) + j_{1}(t)+ j_{2}(t)
\end{equation}
is an oscillating function of time.
For $\Delta \sim J$, as shown in Figs.~\ref{fig:plot2.5} and~\ref{fig:plot3}, we get
\begin{eqnarray}
\nonumber j_{osc}(t) \simeq a_{osc} + e^{-\gamma t/\beta}\big[b \sin((\Delta+V/2) t + \phi)\\
+c \sin((\Delta-V/2) t + \psi)\big]
\label{eq. j oscillations}
\end{eqnarray}
In particular, the oscillations are weak for large $J$ and the decay rate becomes larger as $J/\Delta$ increases.
Furthermore, the stationary current is very low for $V<2\Delta$ while it is large for $V>2\Delta$, in agreement with the conventional result~\cite{Tinkham}.

\begin{figure}
[t!]
\includegraphics[width=0.499\columnwidth]{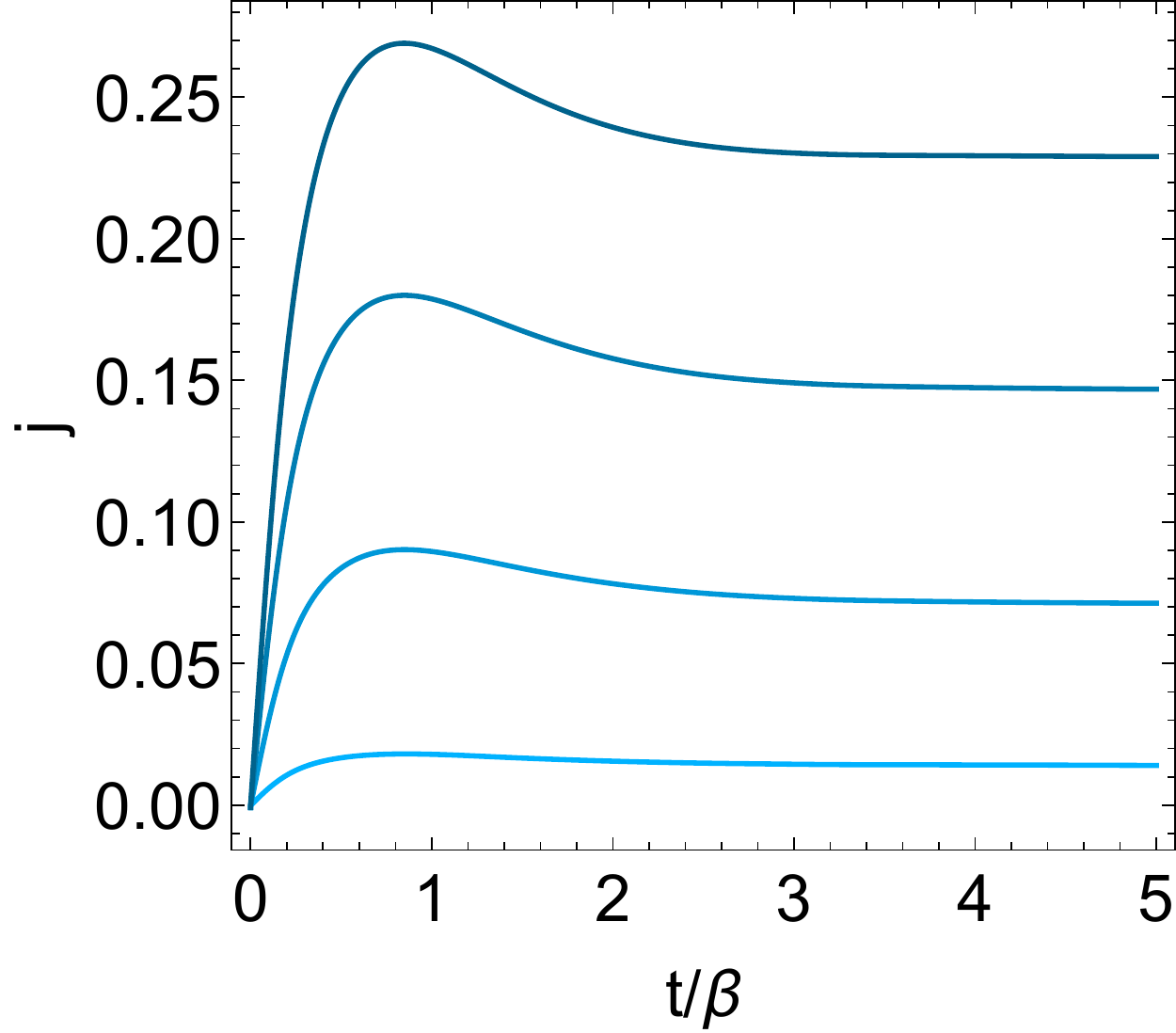}\includegraphics[width=0.499\columnwidth]{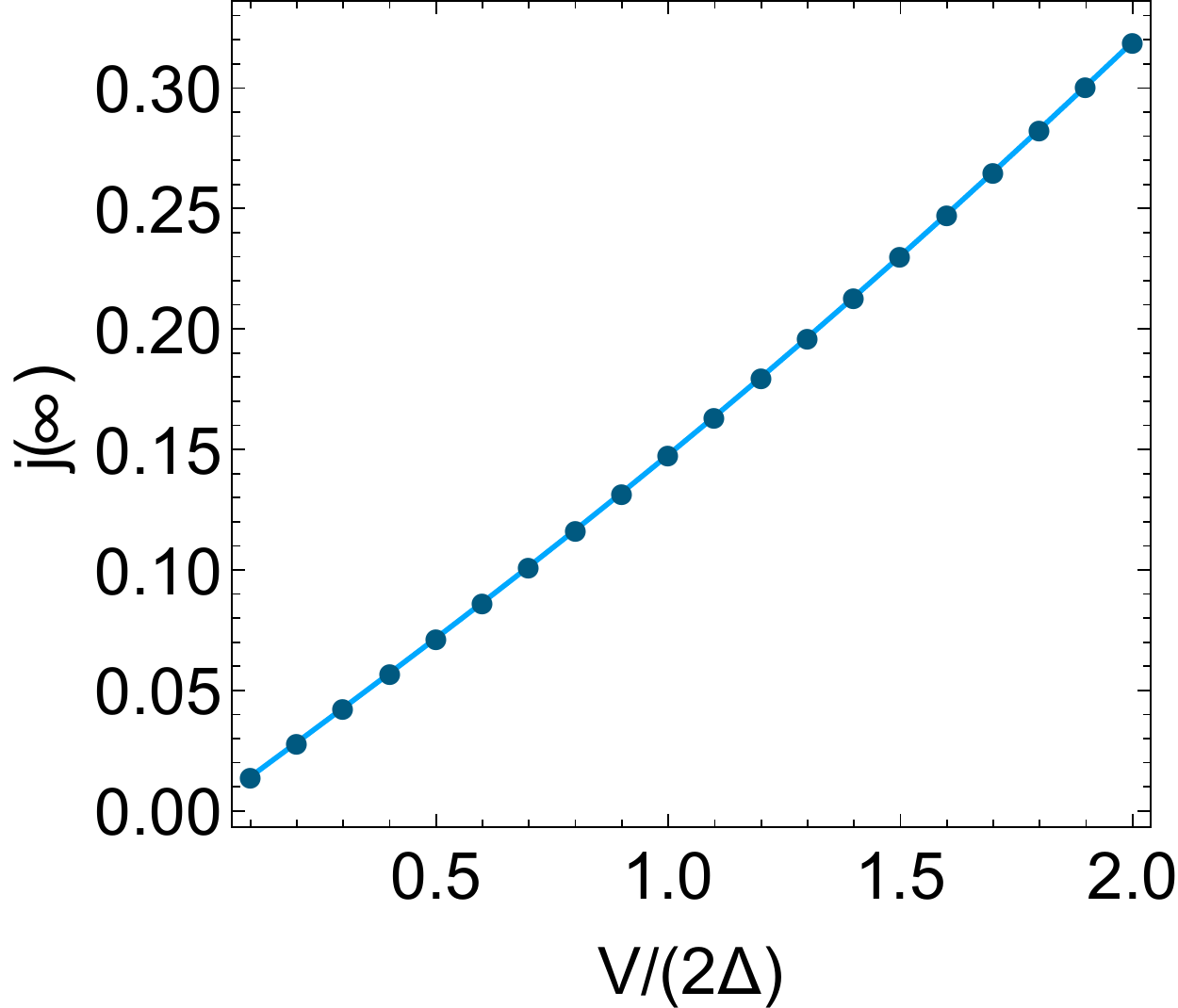}\\
\includegraphics[width=0.499\columnwidth]{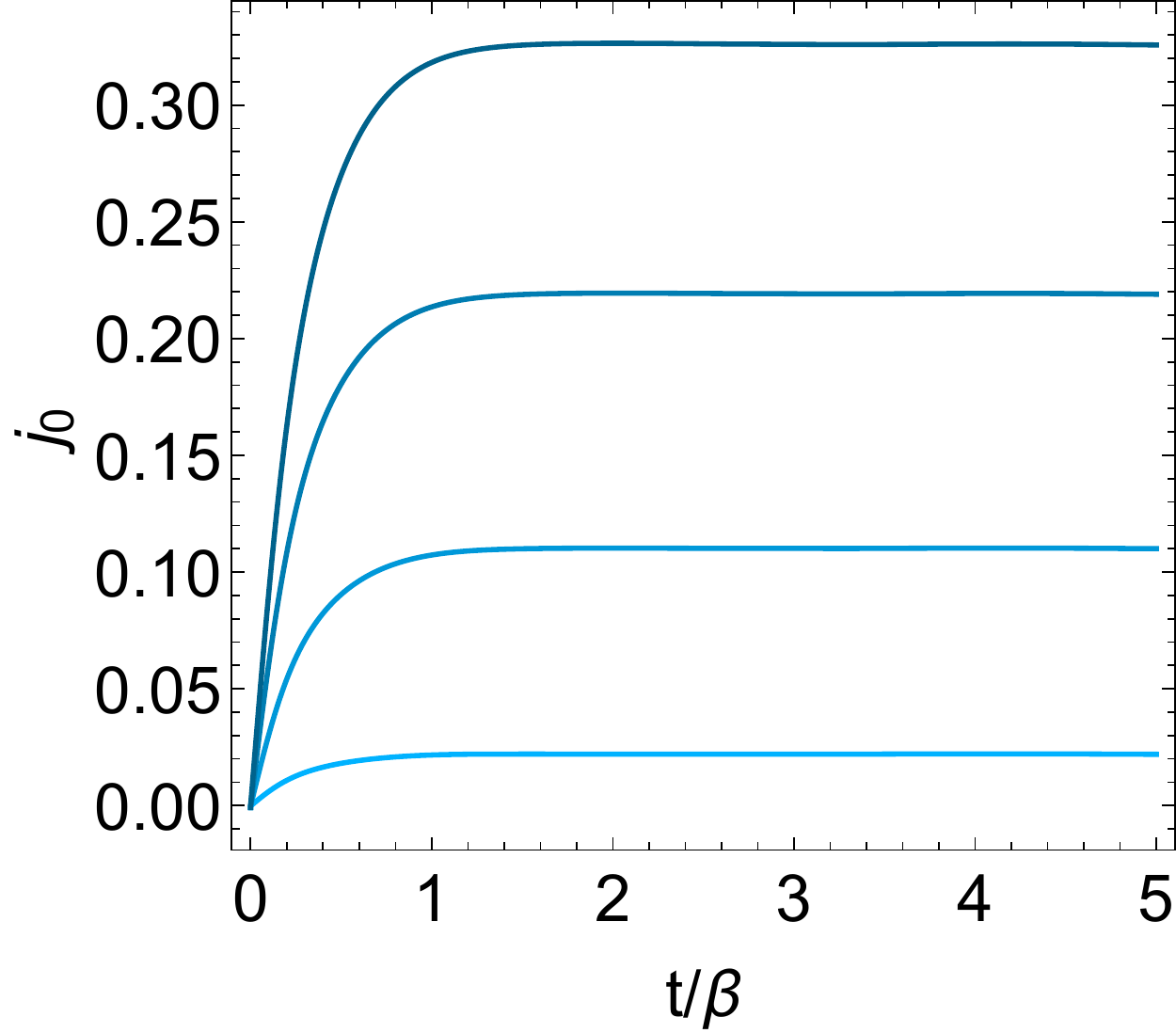}\includegraphics[width=0.499\columnwidth]{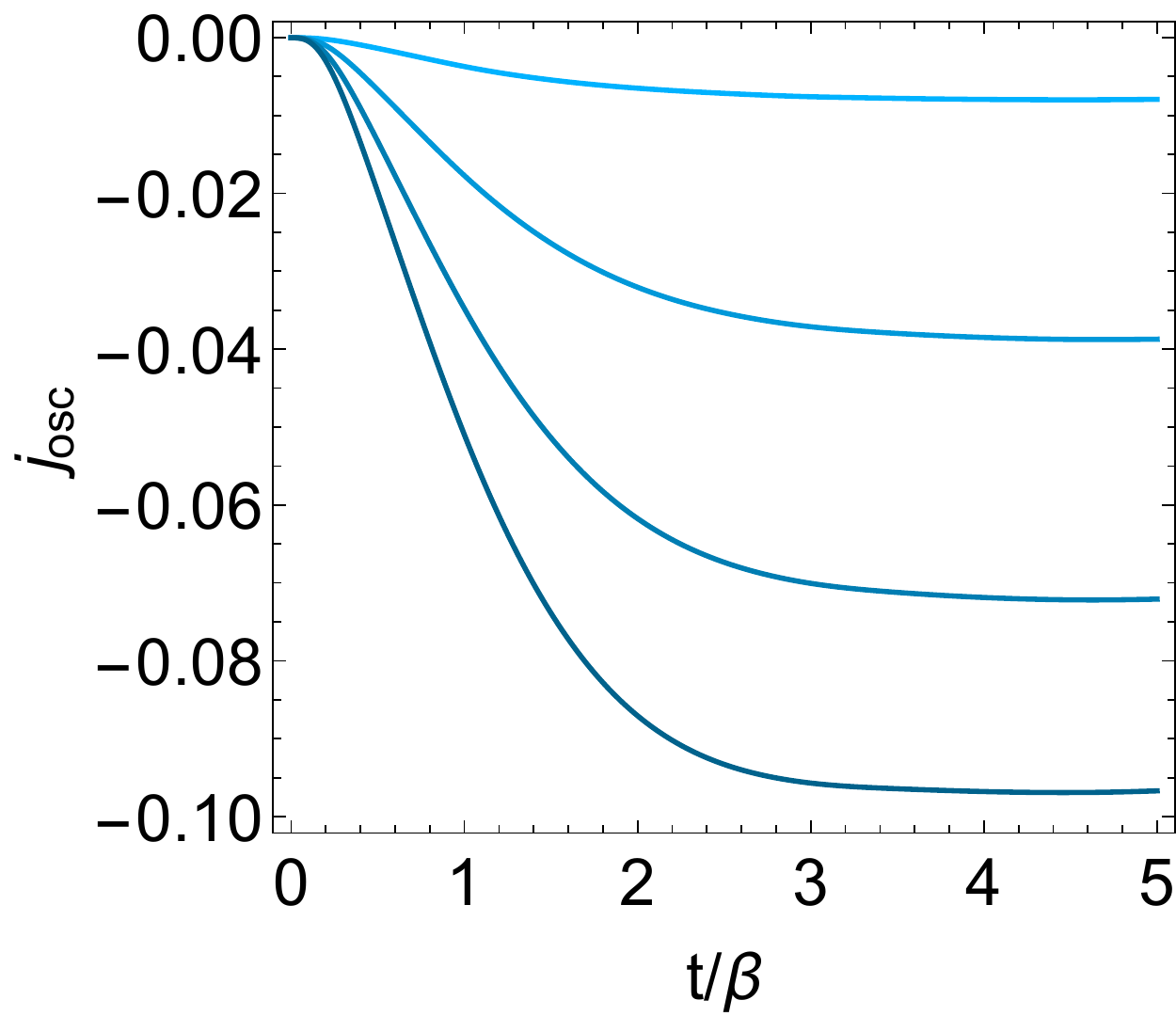}
\caption{{\color{black} SYK$_4$ with superconductive leads:} (Top left panel) The full current $j=j_0+j_{osc}$ as a function of time for different values of $\beta V$. We put $\beta J=10$, $\beta \Delta=0.5$ and $\beta V=0.1,0.5,1,1.5$ (from light to dark lines). (Top right panel) the stationary value of the current $j(\infty)$ calculated averaging over time for $5<t/\beta<10$. (Bottom left time) The normal contribution $j_0$ and (bottom right time) the oscillating contribution $j_{osc}$ to the current as functions of time for $\beta J=10$, $\beta \Delta=0.5$ and $\beta V=0.1,0.5,1,1.5$ (from light to dark lines).}
\label{fig:plot4}
\end{figure}

For this oscillatory contribution $ j_{osc}$ there are three time scales: (i) short times, where $j_{osc}(t)$ starts from zero and rapidly grows until reaches a value of the order of $a_{osc}$, (ii) intermediate times, where $j_{osc}(t)$ shows oscillations, and (iii) large times, where $j_{osc}$ saturates.
For small $\Delta\ll J$, the intermediate time scale (ii) becomes too short and cannot be observed (see Fig.~\ref{fig:plot4}).
Furthermore, since the growing of $j_{osc}$ in the regime (i) is slower than the one of $j_0$ (see Fig.~\ref{fig:plot4}), the full current $j$ reaches a maximum value at short times and then relaxes at later time.
As a function of the voltage, for small $V$, the stationary current is approximately linear in $V$.

\subsection{SYK$_4$ versus SYK$_2$}
\begin{figure}
[t!]
\includegraphics[width=0.499\columnwidth]{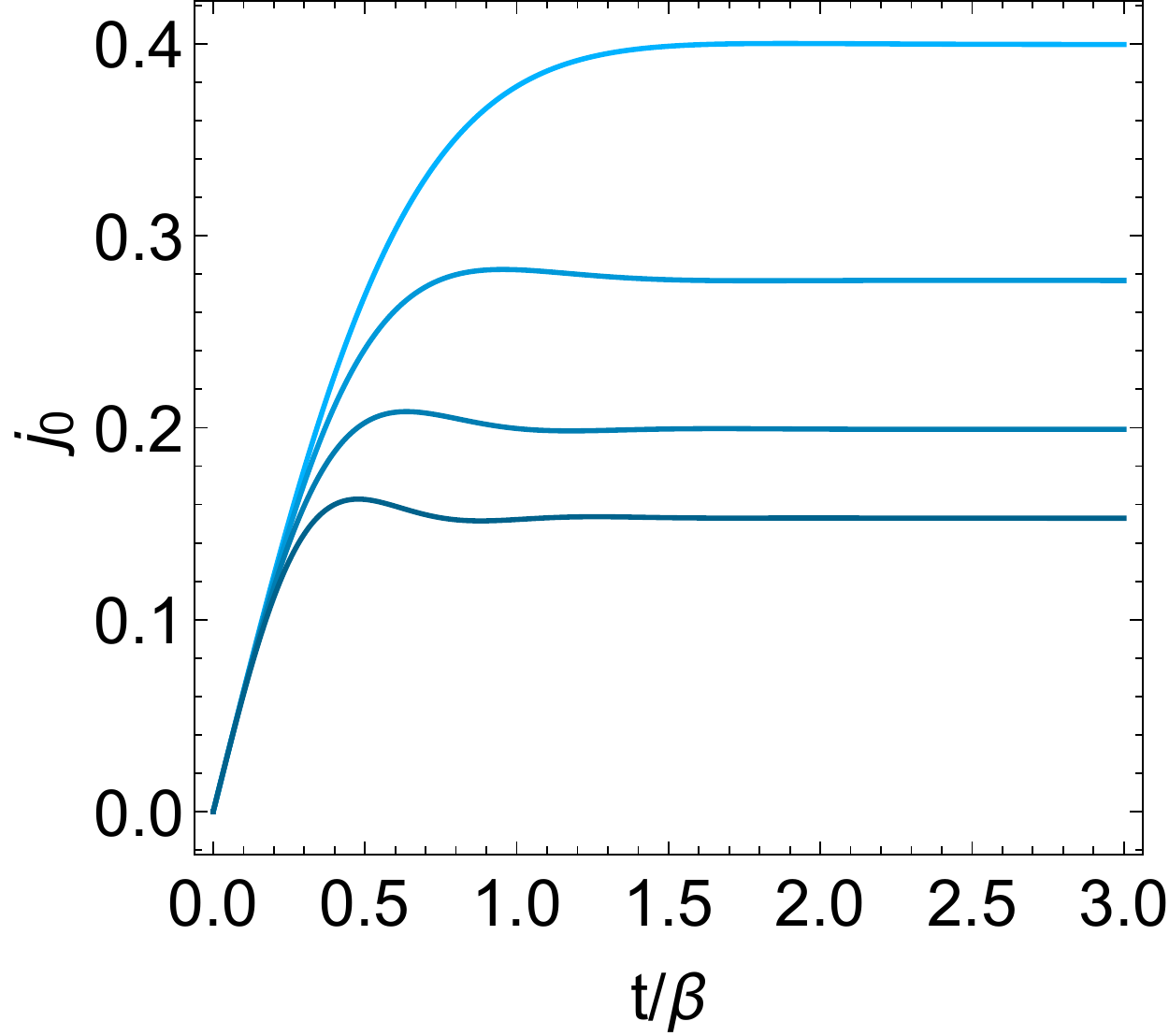}\includegraphics[width=0.499\columnwidth]{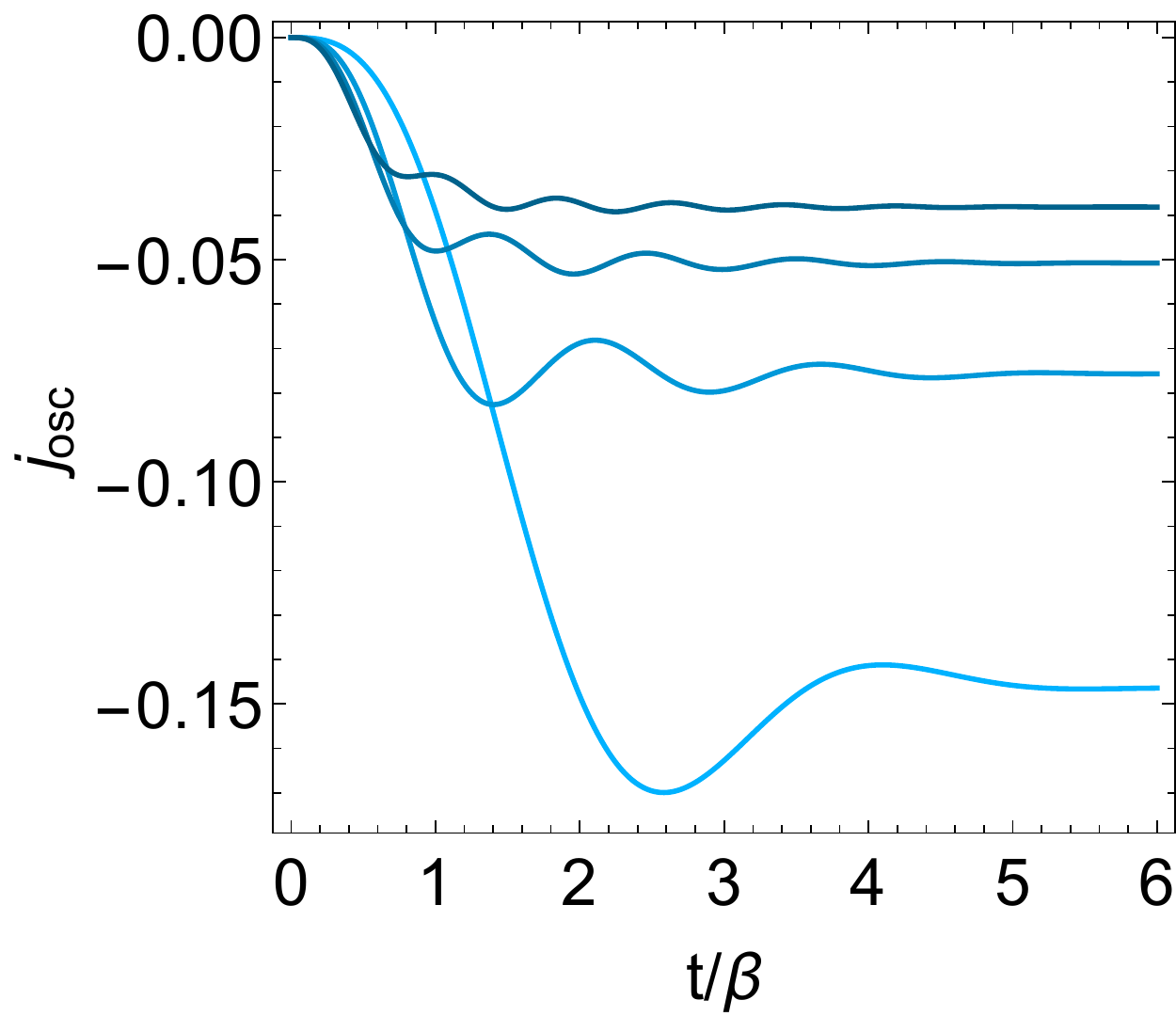}
\caption{{\color{black} SYK$_2$ with superconductive leads:} The current contributions $j_0$ (left panel) and $j_{osc}$ (right panel) as functions of time for different values of $\beta J$ for $q=2$. 
We put $\beta V=1$, $\beta \Delta=0.5$ and $\beta J=1,2,3,4$ (from light to dark lines). }
\label{fig:plot5}
\end{figure}

Finally we are going to compare the cases with $q=4$ and $q=2$, namely SYK$_4$ and SYK$_2$, showing their differences in the current profiles. For $q=2$, and normal leads, $\Delta=0$, the current as a function of time is still a hyperbolic tangent for small $\beta J \ll 1$, see Eq.~(\ref{eq.tanh}). For $\beta J \gtrsim 1 $, the hyperbolic tangent is deformed into a curve exhibiting a maximum at short times (see Fig.~\ref{fig:plot5}). This peak is due to the reaching of a metastable state, which relaxes for larger times. Metastability occurs since for large $J$ there is a separation of the time scales for $q=2$. In contrast for $q=4$ the current monotonically increases with time (for small voltage). For $q=2$, and $\Delta\neq 0$, $j_{osc}$ displays exponentially decaying oscillations. For small $\beta J$, their frequencies are given by $\Delta\pm V/2$, but, increasing $\beta J$, become slower (see Fig.~\ref{fig:plot5}). These oscillations are weak but can be observed also for large $\beta J$. In contrast, for the case with $q=4$, the frequencies of the oscillations are not affected by the coupling $J$ and remain $\Delta \pm V/2$, however they decay fast and are not observed when $J$ becomes large compared to the gap.

\section{Discussion and conclusions}
To give an interpretation in terms of a two-dimensional gravity, we note that, in the holographic limit $1\ll \beta J \ll N $,  the SYK correlation function has the same form of the correlation function of a particle with certain mass, calculated near one of the two boundaries of a two-dimensional anti de-Sitter space-time (AdS$_2$) with metric $ds^2=(-d \tilde t^2 + d z^2)/z^2$. We recall that, for a coordinate $\tilde t(u)$, we get the coordinate $z(u)=\epsilon \tilde t'(u)$, where $u$ is a parameter which corresponds to the time in the SYK model, and $\epsilon$ is small. Thus, we can view our normal transport as the tunneling of the fermions from the left to the right reservoir across a gas of particles near the AdS$_2$ boundary. Initially, every system is in a thermal state with same temperature. When the tunneling coupling is switched on, from the fermions of the left reservoir, new particles are created near the AdS$_2$ boundary. In simple terms, the probability for these particles to fall inside the black hole increases with $u$. However, it is possible to have tunneling with the right reservoir, acquiring a non-zero current.
Since $J/N$ corresponds to $G_N/\bar \phi_r$ where $G_N$ is the gravitational coupling and $\bar \phi_r$ is the renormalized (constant) dilaton field at the boundary, by increasing $J$ more and more particles tend to fall inside the black hole, and the stationary current decreases to zero. Furthermore, from this heuristic point of view, there is no reason to have a peak of the current in time by increasing the coupling $J$. The peak observed for superconducting leads is due to the fact that the tunnelings of $\alpha_\uparrow$ and $\alpha_\downarrow$ are different, producing also the oscillations of the current in time. The frequency of these oscillations is not related to the black hole and is, therefore, independent on $J$.

In conclusion, we have investigated the transport properties across a non-Fermi liquid system described by a SYK model with $q=4$ real fermions. The current as function of time shows an interesting behavior. For normal leads, in the limit of strong coupling and for a small bias, the current increases and reaches monotonically the stationarity, in contrast to the case of a disordered Fermi liquid, e.g., for $q=2$, where metastability occurs. For superconducting leads, for large $\Delta\sim J$, oscillations which decay exponentially having frequencies $\Delta \pm V/2$ are present, however the decay rate becomes large as the gap decreases, therefore, they are strongly suppressed for small gap compared to $J$. In contrast, for $q=2$ the oscillations have frequencies $\Delta \pm V/2$ for small $J$ that smoothly change upon increasing $J$, nevertheless they remain very pronounced even for large $J$. 
{\color{black} Finally, we note that a decay of oscillations of the singlet pair correlation functions following a quantum quench in a SYK model has been recently observed in Ref.~\cite{zhou23}, although in a slightly different system.}

\subsection*{Acknowledgments}
The authors acknowledge financial support from the Project BIRD 2021 ``Correlations, dynamics and topology in long-range quantum systems'' of the Department of Physics and Astronomy, University of Padova,
and from the European Union-Next Generation EU within the ``National Center for HPC, Big Data and Quantum Computing'' (Project No. CN00000013, CN1 Spoke 10 - Quantum Computing).

\appendix

\section{Numerical solution}\label{app.numsol}
We rewrite Eqs.~\eqref{eq.SD1} and~\eqref{eq.SD2} as
\begin{eqnarray}
\label{eq. sp 1} -\partial_t G_0(t,t') + \int_{C'} dt'' \Sigma_0(t,t'')G_0(t'',t') = \delta(t-t')\\
\Sigma_0(t,t') = J^2 G_0^3(t,t')
\end{eqnarray}
We can  numerically solve these equations by dividing the path $C'$ in $N_p$ smaller paths $[t_i,t_{i+1}]$, with $i=1,\ldots, N_p+1$. We integrate Eq.~\eqref{eq. sp 1} in the interval $[t_{i+1},t_{i}]$, thus we get
\begin{equation}\label{eq. discrete}
G_0(t_{i},t_j)- G_0(t_{i+1},t_j) + \int_{t_i}^{t_{i+1}} dt \int_{C'} dt'' \Sigma(t,t'') G_0(t'',t_j)=  I_{ij}
\end{equation}
where $I_{ij}=\delta_{i+1,j}/2+\delta_{i,j}/2$, and by using the trapezoidal rule we get
\begin{eqnarray}
\nonumber &&I_{ij}=G_0(t_{i},t_j)- G_0(t_{i+1},t_j) + \\
\nonumber &&+\, \frac{\delta t_i}{4} \sum_{l=2}^{N_p} \big(\Sigma_0(t_{i+1},t_l) + \Sigma_0(t_i,t_l)\big)G_0(t_l,t_j) (\delta t_l +\delta t_{l-1}) \\
&&+\, \frac{\delta t_i}{4}\big(\Sigma_0(t_{i+1},t_1) + \Sigma_0(t_i,t_1)\big)G_0(t_1,t_j)(\delta t_1+\delta t_{N_p}) 
\label{eq.numeric}
\end{eqnarray}
with $\delta t_j = t_{i+1}-t_i$, and boundary conditions $G_0(t_{N_p+1},t_j)=-G_0(t_1,t_j)$ and $G_0(t_i,t_{N_p+1})=-G_0(t_i,t_1)$, derived from $\chi(t_{t_{N_p+1}})=-\chi(t_1)$. We consider $t_i$ as the Chebyshev nodes, which are
\begin{equation}
t_{i+1} = i\beta/2 +i\beta/2 \cos\frac{(2i-1)\pi}{2 (N_{th}-1)}
\end{equation}
for $i=1,\ldots,N_{th}-1$, and $t_1=i\beta$, $t_{N_{th}+1}=0$,
\begin{equation}
t_{N_{th}+N_K-i+1} = \tau/2 +\tau/2 \cos\frac{(2i-1)\pi}{2 (N_{K}-1)}
\end{equation}
for $i=1,\ldots,N_{K}-1$, and  $t_{N_{th}+ N_{K}+1}=\tau$,
\begin{equation}
t_{N_{th}+N_K+i+1} = \tau/2 +\tau/2 \cos\frac{(2i-1)\pi}{2 (N_{K}-1)}
\end{equation}
for $i=1,\ldots,N_{K}-1$, and  $t_{N_{th}+ 2 N_{K}+1}=0$. We can write Eq.~(\ref{eq.numeric}) in terms of a matrix in time, $M(\Sigma)$, applied to $G_0$, namely we have the equation $M(\Sigma) G_0 = I$ which we have to solve with a self-consistent procedure.

\section{Derivation of the action}\label{app.action}
The SYK action reads
\begin{equation}\label{eq. S}
S_{SYK}= \int_{C'} dt \Big(\frac{i}{2}\sum_{j} \chi_j\dot \chi_j  +\sum_{j<k<l<m} J_{jklm} \chi_j \chi_k \chi_l\chi_m\Big)
\end{equation}
and the tunneling is described by the action
\begin{equation}\label{eq. Stun}
S_T =  \int_{C}dt\Big( \sum_{j,a,k,\sigma} w_{akj}\chi_j \psi_{ak\sigma}\Big)
\end{equation}
where, for brevity, we defined
\begin{equation}
\psi_{ak\sigma} = e^{is_a Vt/2} c^\dagger_{ak\sigma}-e^{-is_a Vt/2} c_{ak\sigma}\,.
\end{equation}
The action $S=S_{SYK}+S_T$ can be expressed as the sum $S=S_0 + \Delta S + S^{(0)}$, where
\begin{eqnarray}
S_0 &=& \int d t \Big(\frac{i}{2} \chi_0\dot \chi_0+\sum_{a,k,\sigma} w_{ak0}\chi_0 \psi_{ak\sigma}\Big) \\
\nonumber S^{(0)} &=& \int dt\Big(\frac{i}{2}\sum_{j>0} \chi_j \dot\chi_j+ \sum_{0<j<k<l<m} J_{jlkm}\chi_j \chi_k \chi_l\chi_m\\
&& +\sum_{j>0,a,k,\sigma} w_{akj}\chi_j \psi_{ak\sigma}\Big)\\
\Delta S &=& \int d t \sum_{0<k<l<m} J_{0lkm}\chi_0 \chi_k \chi_l\chi_m
\end{eqnarray}
We can derive an effective action $S_e$ for the Majorana fermion $j=0$ and the leads by tracing out all the other Majorana fermions (e.g., see Ref.~\cite{Georges96}), such that
\begin{equation}
e^{iS_e} \propto  e^{i S_0} \int \mathcal D \chi e^{i S^{(0)}+i\Delta S}
\end{equation}
where the path integral is over $\chi_j$ with $j>0$. Then
\begin{equation}\label{eq. se}
S_e = S_0 - i \ln \int \mathcal D \chi e^{i S^{(0)}+ i \Delta S}
\end{equation}
up to an irrelevant constant.
We can now calculate the average over the disorder of $S_e$ by means of the replica trick
\begin{equation}
\overline{\ln \int \mathcal D \chi e^{i S^{(0)} +i \Delta S}} = \lim_{M\to 0} \frac{1}{M} \ln \int \mathcal D \chi \overline{e^{ i\sum_{\alpha}(S^{(0)}+\Delta S)}}
\end{equation}
We consider a Gaussian distribution such that $\overline{J^2_{jklm}} = \sigma^2_J$ and $\overline{w^2_{akj}} = \sigma^2_w$ and integrals such as
\begin{equation}
\int \frac{d J_{jklm}}{\sqrt{2\pi} \sigma_J} e^{-\frac{J_{jklm}^2}{2\sigma_J^2}}e^{J_{jklm} x} = e^{\frac{\sigma_J^2 x^2}{2}}
\end{equation}
In order to simplify notation, we define the following quantities, dropping the time dependences,
$\Psi = \sum_{\sigma,\sigma',a,k} \psi_{ak\sigma}(t) \psi_{ak\sigma'}(t')$, $\chi_j^\alpha=\chi_j^\alpha(t)$, $\chi_j^\beta=\chi_j^\beta(t')$, and $\Xi_{\alpha \beta}= \Xi_{\alpha \beta}(t,t') = \sum_{j>0} \chi_j^\alpha(t) \chi_j^\beta(t')/N$. We have, then,
\begin{eqnarray}
&&\nonumber \hspace{-0.2cm}X\equiv \int \mathcal D \chi \overline{e^{i\sum_{\alpha}(S^{(0)}+\Delta S)}} \\
&&\nonumber = \int \mathcal D \chi \mathcal D J  \mathcal D w e^{-\sum_{j<k<l<m}\frac{J_{jklm}^2}{2\sigma_J^2}} e^{-\frac{\sum_{a,j>0,k}w_{akj}^2}{2\sigma_w^2}} e^{i\sum_{\alpha}(S^{(0)}+\Delta S)}\\
&&\nonumber  = \int \mathcal D \chi \exp\bigg\{ - \int dt \frac{1}{2}\sum_{j>0,\alpha}  \chi_j^\alpha \dot\chi_j^\alpha \\
&&\nonumber  -\, \frac{\sigma_J^2N^3}{2(3!)}\sum_{\alpha,\beta}\int dt dt' \left( \chi^\alpha_0 \chi^\beta_0 \Xi_{\alpha \beta}^3
+ \frac{N}{4} \Xi_{\alpha \beta}^4\right) \\
&&+\, \frac{\sigma_w^2}{2}\sum_{\alpha,\beta}\int dt dt' \left( N \Xi_{\alpha \beta} \Psi \right)\bigg\}\;
\end{eqnarray}
We consider $\sigma_J^2 = 3! J^2/N^3$. For a given a function $f(x)$, we can use the following identity
\begin{equation}
f(\Xi) = \int dx f(x) \delta(x-\Xi) = \frac{N}{2\pi} \int dx dy f(x) e^{iNy(x-\Xi)}
\end{equation}
Considering $x=-G_{\alpha \beta}(\tau,\tau')$, $y=-i \Sigma_{\alpha\beta}(\tau,\tau')/2$ and the functions $f(\Xi) = e^{-J^2 \chi^\alpha_0 \chi^\beta_0 \Xi^3}$ and $f(\Xi) = e^{-J^2 N \Xi^4/4}$, we get
\begin{eqnarray}
\nonumber X = \int \mathcal D G \mathcal D \Sigma \mathcal D \chi \exp\bigg\{ - \int dt \frac{1}{2}\sum_{j>0,\alpha}  \chi_j^\alpha \dot\chi_j^\alpha \\
\nonumber +\, \frac{1}{2}\sum_{\alpha,\beta}\int dt dt'\bigg( J^2 \chi^\alpha_0 \chi^\beta_0G_{\alpha\beta}^3 - N \frac{J^2}{4} G_{\alpha\beta}^4 \\
+\, N\Sigma_{\alpha\beta}(G_{\alpha\beta}+\Xi_{\alpha\beta})+N\sigma_w^2 \Xi_{\alpha\beta}\Psi\bigg) \bigg\}
\end{eqnarray}
the integral over the Grassmann variables is $(N-1)\ln(\text{Pf}(-\partial_t+\Sigma + \sigma_w^2\Psi ))$, since $\sigma_{w}^2 \sim 1/N$, we get
\begin{eqnarray}
\nonumber && (N-1)\ln(\text{Pf}(-\partial_t+\Sigma  + \sigma_w^2\Psi ))\sim(N-1)\ln(\text{Pf}(-\partial_t+\Sigma)) \\
&&+ \frac{N}{2}\Tr{(-\partial_t+\Sigma)^{-1} \sigma_w^2\Psi}
\end{eqnarray}
then
\begin{eqnarray}
&& \nonumber X = \int \mathcal D G \mathcal D \Sigma  \exp\Bigg\{(N-1)\ln(\text{Pf}(-\partial_t+\Sigma))\\
&&+\, \frac{N}{2}\int dt dt'{(-\partial_t+\Sigma)^{-1} \sigma_w^2\Psi} \\
&&\nonumber + \,\frac{1}{2} \sum_{\alpha,\beta}\int dt dt'\left( J^2 \chi^\alpha_0 \chi^\beta_0G_{\alpha\beta}^3 - N \frac{J^2}{4} G_{\alpha\beta}^4 + N\Sigma_{\alpha\beta}G_{\alpha\beta}\right) \Bigg\}
\end{eqnarray}
for $N\to\infty$, we consider the saddle-point of the function which goes as $N$, which is solution of the equations
\begin{eqnarray}
\Sigma_{\alpha\beta}=J^2 G^3_{\alpha\beta}\\
\hat G=(-\partial_\tau+\hat\Sigma)^{-1}
\end{eqnarray}
where $\hat G$ is the matrix with elements $G_{\alpha,\beta}$.
We consider $\Sigma_{\alpha \beta}(t,t') = \delta_{\alpha,\beta}\Sigma_0(t,t') $ and $G_{\alpha \beta}(t,t') = \delta_{\alpha,\beta}G_0(t,t') $, then
\begin{equation}
X\propto e^{\frac{M}{2}\int dt dt' \chi_0(t) \chi_0(t') \Sigma_0(t,t')+\frac{MN\sigma_w^2}{2} \int dt dt' G_0(t,t') \Psi(t,t')}
\end{equation}
so that we get Eq.~\eqref{eq. S eff}.

\section{Derivation of the average current}\label{app.avecur}

We aim to perform the average over disorder of the current in Eq.~\eqref{eq. JL cos 11 0}. Let us start to focus on the sum with $j=j'$, i.e., on the term $\overline{J}_{L,d}(t)$. 
We can calculate the average of $G_{jj}(t,t_1)$  over all the random variables except $\{w_{akj}\}_{a,k}$, namely, except those with site index $j$, by using an effective action for the $j$-th fermion and the leads. This action can be derived by using the dynamical mean field theory for $N\to \infty$ and reads (see Appendix \ref{app.action})
\begin{eqnarray}\label{eq. S eff}
\nonumber S&=&\int_{C'} dt \frac{i\chi(t) \dot\chi(t)}{2} - \frac{i}{2}\int_{C'} \int_{C'} dt dt' \chi(t)\chi(t') \Sigma_0(t,t')\\
 &&+ S'_L+S'_R+S'_T
\end{eqnarray}
where
\begin{equation}
S'_T= \int_C d t \sum_{a,k,\sigma} w_{akj}\chi \psi_{ak\sigma}
\end{equation}
and
\begin{equation}\label{eq. leads action}
S'_L+S'_R= S_L+S_R - \frac{i}{2}\int_{C} \int_{C} dt dt' N \sigma_w^2 G_0(t,t') \Psi(t,t')
\end{equation}
where $S_a$, with $a=L, R$, are the actions of the free leads, and $\Psi(t,t') = \sum_{\sigma,\sigma',a,k} \psi_{ak\sigma}(t) \psi_{ak\sigma'}(t')$, where
\begin{equation}
\psi_{ak\sigma} = e^{is_a Vt/2} c^\dagger_{ak\sigma}-e^{-is_a Vt/2} c_{ak\sigma}
\end{equation}
if $t\in C$ and $\psi_{ak\sigma}=0$ otherwise. Thus, the actions of the leads are modified into $S'_a$ due to the coupling with the SYK model.
By performing the average over the disorder of the current in Eq.~\eqref{eq. JL cos 11 0}, we get
\begin{eqnarray}\label{eq. JL cos 11 2}
\nonumber \overline{J}_{L,d}(t) &=& 2\text{Re} \sum_{k,j} \int  \prod_{a,k}\mathcal Dw_{akj}  e^{-\frac{\sum_{a,k}w_{akj}^2}{2\sigma_w^2}} w_{Lkj}^2 e^{iVt/2}\\
\nonumber && \times i\int_Cdt_1e^{-iVt_1/2}\big((1 +\cos 2\theta_k)g_{Lk\sigma}(t_1,t)\\
 && -(1-\cos 2\theta_k)g_{Lk\sigma}(t,t_1)\big)G(t,t_1)
\end{eqnarray}
where $G(t,t_1)$ is the average of $G_{jj}(t,t_1)$ over all the parameter except $\{w_{akj}\}_{a,k}$,
calculated by using the effective action in Eq.~\eqref{eq. S eff}. $G(t,t')$ is then the solution of the following Dyson equation
\begin{equation}\label{eq. Dyson}
G(t,t') = G_0(t,t') + \int dt_1 dt_2 G_0(t,t_1) \Sigma'(t_1,t_2) G(t_2,t')
\end{equation}
where we define the self-energy
\begin{equation}
\Sigma'(t_1,t_2) = \sum_{a,k,\sigma} w_{akj}^2 e^{-i s_a(t_1-t_2)V/2} g'_{ak\sigma}(t_1,t_2)
\end{equation}
where the Keldysh Green function $g'_{ak\sigma}(t_1,t_2)$ are calculated with respect to the action of the leads in Eq.~\eqref{eq. leads action}. The  solution $G(t,t')$ of the Dyson equation formally is a power series of the variables $\{w_{akj}^2 \}_{a,k}$, therefore, we can easily calculate the integral over $w_{akj}$ in Eq.~\eqref{eq. JL cos 11 0}, since the distribution is Gaussian, getting
Eq.~\eqref{eq. JL cos}, 
where $G(t,t_1)$ is still obtained by using the effective action in Eq.~\eqref{eq. S eff}, i.e., by solving the Dyson equation in Eq.~\eqref{eq. Dyson}, but with $\sigma_w$ instead of $w_{akj}$. {\color{black} In particular, since $\sigma_w^2\sim 1/N$, from the Dyson equation in Eq.~\eqref{eq. Dyson} we get $G(t,t')=G_0(t,t')+O(1/N)$.

On the other hand, 
concerning the term $\overline{J}_{L,od}(t)$, we can proceed similarly, writing a Dyson equation for $G_{jj'}$ 
for $j\neq j'$, 
\begin{eqnarray}\label{eq. Dyson two sites}
\nonumber && G_{jj'}(t,t')
= G_{0,jj'}(t,t')\\
\nonumber && + \int dt_1 dt_2 G_{0,jj}(t,t_1) \Sigma'_{jj'}(t_1,t_2) G_{j'j'}(t_2,t')\\
\nonumber && 
+  \int dt_1 dt_2 G_{0,jj}(t,t_1) \Sigma'_{jj}(t_1,t_2) G_{jj'}(t_2,t')\\
\nonumber&&+  \int dt_1 dt_2 G_{0,jj'}(t,t_1) \Sigma'_{j'j'}(t_1,t_2) G_{j'j'}(t_2,t')\\
&&+  \int dt_1 dt_2 G_{0,jj'}(t,t_1) \Sigma'_{j'j}(t_1,t_2) G_{jj'}(t_2,t')\,,
\end{eqnarray}
where the self-energy goes as $\Sigma'_{jj'}(t_1,t_2)\sim w_{j} w_{j'}$, so that only the first two terms in the right hand side might give leading terms of order $O(1/N)$. The average of the first term of the current coming from $ G_{0,jj'}$ is trivially zero, so that the contribution to the current comes from the second term. We have to calculate, therefore, the average
\begin{eqnarray}\label{eq.calcolo}
\nonumber &&\int  \prod_{a,k}\mathcal D w_{akj}\mathcal D w_{akj'}  e^{-\frac{\sum_{a,k}(w_{akj}^2+w_{akj'}^2)}{2\sigma_w^2}} w_{Lkj}w_{Lkj'} G_{jj'}(t,t_1) \\
\nonumber && =\int  \prod_{a,k}\mathcal D w_{akj}\mathcal D w_{akj'}  e^{-\frac{\sum_{a,k}(w_{akj}^2+w_{akj'}^2)}{2\sigma_w^2}} w_{Lkj}w_{Lkj'}\\
\nonumber && \times \int dt'_1 dt_2 G_{0,jj}(t,t'_1) \Sigma'_{jj'}(t'_1,t_2) G_{j'j'}(t_2,t')\\
\nonumber&& = \sigma^4_w \int dt'_1dt_2 G_0(t,t'_1) e^{-i(t'_1-t_2)\frac{V}{2}} \sum_{\sigma'}g'_{L k \sigma'}(t'_1,t_2)G_0(t_2,t_1)\\
&&\equiv \sigma^4_w \,\widetilde{G}(t,t_1)
\end{eqnarray}
where we considered the self-energy
\begin{equation}
\Sigma'_{jj'}(t_1,t_2) = \sum_{a,k,\sigma} w_{akj}w_{akj'} e^{-i s_a(t_1-t_2)V/2} g'_{ak\sigma}(t_1,t_2)\,,
\end{equation}
As a result, we get the off-diagonal contribution to the current 
given by Eq.~\eqref{eq. JL cos 11 off-d}. 

In summary, the off-diagonal contribution to the current comes from averaging $w_jw_{j'}G_{jj'}$. 
On the contrary, if we look, instead, only to the averaged Green's functions}, the 
 off-diagonal terms $G_{jj'}$, with $j\neq j'$, eventually generated from the coupling to the leads by multiple tunneling processes are suppressed by  disorder averaging. This can be seen also diagrammatically since, after expanding the Dyson equation, only terms proportional to products of $w_j^{2n}$ survive the averaging over Gaussian disorder, replacing those parameters by $\sigma_w^{2n}$, while the bare Green's functions are independent on the fermionic indices. As a result all the terms in the diagrammatic expansion are diagonal and uniform so that the index $j$ can be safely dropped out.

{\color{black}
\section{Constant density of states} \label{app.green}
To calculate the current $\overline{J}_{L,d}$ in Eq.~(\ref{eq. JL cos}) we have to evaluate the sum
\begin{equation}
\sum_k (1+\cos(2\theta_k)) g_{Lk\sigma}(t_1,t) - (1-\cos(2\theta_k)) g_{Lk\sigma}(t,t_1)
\end{equation}
where $t$ belongs to the backward branch. Let us show that if $\rho(\xi)=\rho$ constant, the contribution coming from the terms with $\cos(2\theta_k)$ is zero, namely we get
\begin{equation}
g_+\equiv \sum_k \cos(2\theta_k) (g_{Lk\sigma}(t_1,t) + g_{Lk\sigma}(t,t_1))=0
\end{equation}
We consider $t_1$ belonging to the backward branch. We get
\begin{eqnarray}
g_+ &=& \sum_k \cos(2\theta_k) (g^{\tilde T}_{Lk\sigma}(t_1,t) + g^{\tilde T}_{Lk\sigma}(t,t_1))\\
 \nonumber &=& i \sum_k \frac{\sqrt{\Delta^2-\lambda_k^2}}{|\lambda_k|} \left(2\frac{\cos(\lambda_k(t_1-t))}{1+e^{\beta \lambda_k}}- e^{i\lambda_k|t_1-t|}\right)
\end{eqnarray}
where we noted that $\cos(2\theta_k) = \xi_k /\lambda_k = \sqrt{\Delta^2-\lambda_k^2}/|\lambda_k|$. In the continuum limit, we get
\begin{equation}
\sum_k \dots = \left(\frac{L}{\pi}\right)^d \int \rho (\xi) d\xi \dots
\end{equation}
which, in constant density of states approximation reads
\begin{equation}
\left(\frac{L}{\pi}\right)^d \rho \int  d\xi \dots=  \left(\frac{L}{\pi}\right)^d \rho \int_I  \frac{|\lambda|}{\sqrt{\Delta^2-\lambda^2}}d\lambda \dots
\end{equation}
where $I=(-\infty,-\Delta]\cup [\Delta,\infty)$. We have, therefore,
\begin{equation}
g_+ \propto \int_I d\lambda \left(2\frac{\cos(\lambda(t_1-t))}{1+e^{\beta \lambda}}- e^{i\lambda|t_1-t|}\right)=0
\end{equation}
since
\begin{eqnarray}
&& \int_I d\lambda \left(2\frac{\cos(\lambda(t_1-t))}{1+e^{\beta \lambda}}- e^{i\lambda|t_1-t|}\right) \\
\nonumber &&= \int_{\Delta}^\infty 2 \cos(\lambda(t_1-t)) \left( \frac{1}{1+e^{\beta \lambda}}+\frac{1}{1+e^{-\beta \lambda}}-1\right)
\end{eqnarray}
which is zero, being $\frac{1}{1+e^{\beta \lambda}}+\frac{1}{1+e^{-\beta \lambda}}-1=0$.
The same result holds if  $t_1$ belongs to the forward branch.
}


\begin{thebibliography}{99}

\bibitem{SachdevYe} S. Sachdev and J. Ye, Phys. Rev. Lett. 70, 3339 (1993).

\bibitem{kitaevtalk} A. Kitaev, ``A simple model of quantum holography.'' http://online.kitp.ucsb.edu/online/entangled15/kitaev/, http://online.kitp.ucsb.edu/online/entangled15/kitaev2/. Talks at KITP, April 7, 2015 and May 27, 2015.

\bibitem{Chowdhury22} D. Chowdhury, A. Georges, O. Parcollet, and S. Sachdev, Rev. Mod. Phys. 94, 035004 (2022).

\bibitem{Patel19} A. A. Patel, S. Sachdev, Phys. Rev. Lett. 123, 066601 (2019).

\bibitem{Sachdev23} S. Sachdev, arXiv:2305.01001 (2023).

\bibitem{kitaev18} A. Kitaev, S. J. Suh, JHEP 05, 183 (2018).

\bibitem{maldacena16} J. Maldacena, D. Stanford, Phys. Rev. D 94, 106002 (2016).

\bibitem{sarosi19} G. S\'{a}rosi, arXiv:1711.08482 (2019).

\bibitem{Franz18} M. Franz, M. Rozali, Nature Reviews Materials 3, 491–501 (2018).

\bibitem{hauke}  P. Uhrich, S. Bandyopadhyay, N. Sauerwein, J. Sonner, J-P. Brantut, P. Hauke, arXiv:2303.11343.

\bibitem{song17} X.-Y. Song, C.-M. Jian, and L. Balents, Phys. Rev. Lett. 119, 216601 (2017).
\bibitem{davison} R. A. Davison, W. Fu, A. Georges, Y. Gu, K. Jensen, and S. Sachdev, Phys. Rev. B 95, 155131 (2017)
\bibitem{Gnezdilov18} N. V. Gnezdilov, J. A. Hutasoit, and C. W. J. Beenakker, Phys. Rev. B 98, 081413(R) (2018).
\bibitem{Can19} O. Can, E. M. Nica, and M. Franz, Phys. Rev. B 99, 045419 (2019).
\bibitem{Guo20} H. Guo, Y. Gu, S. Sachdev, Annals of Physics 418, 168202 (2020).
\bibitem{Kulkarni22} A. Kulkarni, T. Numasawa, and S. Ryu, Phys. Rev. B 106, 075138 (2022).



\bibitem{Almheiri19} A. Almheiri, A. Milekhin, and B. Swingle, arXiv:1912.04912.
\bibitem{Zhang19} P. Zhang, Phys. Rev. B 100, 245104 (2019).

\bibitem{Cheipesh21} Y. Cheipesh, A. I. Pavlov, V. Ohanesjan, K. Schalm, and N. V. Gnezdilov, Phys. Rev. B 104, 115134 (2021).

\bibitem{Eberlein17} A. Eberlein, V. Kasper, S. Sachdev, and J. Steinberg, Phys. Rev. B 96, 205123 (2017).


\bibitem{Zhou20} T.-G. Zhou, and P. Zhang, Phys. Rev. B 102, 224305 (2020).
\bibitem{Kruchkov20} A. Kruchkov, A. A. Patel, P. Kim, and S. Sachdev, Phys. Rev. B 101, 205148 (2020).
\bibitem{Altland19} A. Altland, D. Bagrets, and A. Kamenev, Phys. Rev. Lett. 123, 226801 (2019).


\bibitem{Jacquet20} R. Jacquet, A. Popoff, K.-I. Imura, J. Rech, T. Jonckheere, L. Raymond, A. Zazunov, T. Martin, Phys. Rev. B 102, 064510 (2020).


\bibitem{Macieszczak21} K. Macieszczak, D. C. Rose, I. Lesanovsky, and J. P. Garrahan, Phys. Rev. Research 3, 033047 (2021).



\bibitem{Jauho94} A.-P. Jauho, N. S. Wingreen, and Y. Meir, Phys. Rev. B 50, 5528 (1994).



\bibitem{Tinkham} M. Tinkham, Introduction to superconductivity, 2nd ed. (McGraw-Hill, New York, 1975).


\bibitem{zhou23} T.-G. Zhou, P. Zhang, SciPost Phys. 15, 108 (2023). 

\bibitem{Georges96} A. Georges, G. Kotliar, W. Krauth, and M. J. Rozenberg, Rev. Mod. Phys. 68, 13 (1996).

\end{thebibliography}
\end{document}